\documentclass[
 reprint,
nofootinbib,
 amsmath,amssymb,
 aps,
showkeys
]{revtex4-1}

\usepackage{graphicx}
\usepackage{dcolumn}
\usepackage{bm}
\usepackage{hyperref}
\usepackage{multirow}
\usepackage[dvipsnames]{xcolor}
\hypersetup{
    colorlinks=true,
    linkcolor=MidnightBlue,
    citecolor=MidnightBlue,
    filecolor=MidnightBlue,      
    urlcolor=MidnightBlue
    }
\usepackage[nameinlink,capitalise]{cleveref}
\usepackage{enumitem}

\newcommand{\gsf}{{\sc gsf}}
\newcommand{\sibyll}[1]{{\sc sibyll#1}}
\newcommand{\ddm}{{\sc ddm}}
\newcommand{\mceq}{{\sc mceq}}

\newcommand{\proposal}{{\sc proposal}}
\newcommand{\mute}{{\sc mute}}
\newcommand{\daemonflux}{{\sc daemonflux}}
\newcommand{\mum}{{\sc mum}}
\newcommand{\music}{{\sc music}}
\newcommand{\musun}{{\sc musun}}
\newcommand{\geant}{{\sc Geant4}}

\begin{document}


\title{Cosmic ray muons in laboratories deep underground}

\author{William Woodley}
\email{wwoodley@ualberta.ca}
\affiliation{University of Alberta, Department of Physics,\\
4-181 CCIS, 116 St.\ and 85 Ave., Edmonton, AB T6G 2R3, Canada}

\author{Anatoli Fedynitch}
\email{anatoli@gate.sinica.edu.tw}
\affiliation{Institute of Physics, Academia Sinica, Taipei City, 11529, Taiwan}

\author{Marie-C\'ecile Piro}
\email{mariecci@ualberta.ca}
\affiliation{University of Alberta, Department of Physics\\
4-181 CCIS, 116 St.\ and 85 Ave., Edmonton, AB T6G 2R3, Canada}

\date{\today}

\begin{abstract}
We provide comprehensive calculations of total muon fluxes, energy and angular spectra, and mean muon energies in deep underground laboratories---under flat overburdens and mountains and underwater---using our latest calculation code, \mute{} v3. For precise modeling, we compiled rock densities and chemical compositions for various underground labs, as well as topographic map profiles of overburdens, and integrated them into our calculations. Our results show excellent agreement with available data for most underground sites when using the latest surface muon flux model, \daemonflux{}. Moreover, since our calculations do not rely on underground measurements of muons or other secondaries, we can verify the consistency of measurements across different detectors at different sites. \mute{} is an open-source, publicly available program, providing a solid framework for accurate muon flux predictions in various underground environments, essential for applications in cosmic ray physics and dark matter searches.
\end{abstract}


\maketitle

\section{\label{sec:introduction}Introduction}

For many decades, measurements of muons have been conducted underwater and underground to study the energy spectrum and composition of primary cosmic rays \cite{Barrett:1952woo,ParticleDataGroup:2022pth}. Despite these long-standing efforts, large uncertainties surrounding atmospheric muon and high-energy neutrino fluxes persist, with sustained challenges to their reduction~\cite{Barr:2006it,Honda:2006qj,Yanez:2023lsy}. Shielded by the Earth's natural overburden, large-volume underground detectors offer a unique opportunity to study high-energy muons without the technical challenges or high cost of single-purpose surface-based spectrometers capable of resolving multihundred TeV range energies.

Addressing these uncertainties is of paramount importance for several reasons. Precise estimation of the muon flux can significantly benefit fields like rare-event searches, including direct dark matter searches. The detectors for these searches are typically installed deep underground in mines or beneath mountains in order to limit exposure to cosmic-ray muons~\cite{Formaggio:2004mm}. However, cosmic-ray muons produce secondary neutrons by interacting with the rock or surrounding detector materials, which pose significant identification challenges, as they can mimic low-energy dark matter signals. Knowledge of the underground muon rate, energy spectrum, and angular distribution is necessary to estimate muon-induced backgrounds to design effective shielding, including muon and neutron vetos. For the past few decades, analytical calculations~\cite{Bugaev:1998bi,Lipari:1991ut} and empirical parametric fits to vertical-equivalent muon intensity data~\cite{crouch1987improved,Mei:2005gm} have served as inputs to Monte Carlo muon transport codes such as \mum{}~\cite{Sokalski:2000nb}, \music{} and \musun{}~\cite{Kudryavtsev:2008qh}, and \geant{}~\cite{GEANT4:2002zbu}. However, these methods often demand significant computing resources and do not thoroughly address the treatment of errors. \textsc{fluka}~\cite{Battistoni:2015epi}, meanwhile, can be used to calculate muon fluxes starting from a cosmic ray flux parametrization; however, it also comes with computational cost, in particular for high energies.

In our previous work in Ref.~\cite{Fedynitch:2021ima}, we concentrated on the development of the open-source \mute{} code, designed to calculate underground muon fluxes and other related observables without using those fluxes as input. \mute{} is flexible, efficient, and precise, and combines two modern computational tools, \mceq{}~\cite{Fedynitch:2015kcn,Fedynitch:2018cbl} and \proposal{}~\cite{koehne2013proposal}, for surface muon fluxes and underground transport respectively. We previously compared vertical-equivalent underground intensities computed with \mute{} using a flat overburden to experimental measurements. In the present study, we expand on this by scrutinizing the residual body of related data using an updated version of \mute{} in combination with the latest and most precise surface flux model, the \daemonflux{} model~\cite{Yanez:2023lsy}. \daemonflux{} makes use of two data-driven models to achieve uncertainties on surface fluxes of less than 10\% up to 1 TeV: Global Spline Fit (\gsf{})~\cite{Dembinski:2017zsh} for the primary cosmic ray flux, and the Data-Driven Model (\ddm{})~\cite{Fedynitch:2022vty} for hadronic interactions. In this work, we calculate total underground and underwater muon fluxes, as well as underground angular distributions, energy spectra, and mean energies. We use topographic maps for laboratories situated under mountains and we incorporate the density and composition of the rock above each lab into our calculations, which are significant sources of systematic uncertainty. We compare our \mute{} predictions with experimental data from different underground and underwater sites worldwide. The new release of \mute{} is publicly available so our results can be easily reproduced and used for further studies.\footnote{\url{https://github.com/wjwoodley/mute}}

\section{\label{sec:computationalmethod}Computational Method}

For practical applications, the muon intensity at a certain depth underground has been historically parametrized using functions termed depth-intensity relations (DIR). D.-M.\ Mei and A.\ Hime~\cite{Mei:2005gm} give the following DIR for depth ranges between 1 and 10~km.w.e.:

\begin{equation}
\label{eq:DIR_PDG}
I(h)=I_1e^{(-h/\lambda_1)}+I_2e^{(-h/\lambda_2)},
\end{equation}

\noindent where $h$ is the vertical depth of the lab, and $I_1$, $I_2$, $\lambda_1$, and $\lambda_2$ are determined by fitting to experimental data. Comparatively, many collaborations~\cite{FREJUS:1989lko,MACRO:1990ykz,MACRO:1995egd,LVD:1998lir,SNO:2009oor} give the following three-parameter relation:

\begin{equation}
\label{eq:DIR_LVD}
I(h)=A\left(\frac{h_0}{h}\right)^{\alpha}e^{-h/h_0},
\end{equation}

\noindent where $A$, $h_0$, and $\alpha$ are fit from data. Parametric fits like these can provide reasonable estimates for underground labs, as demonstrated in Ref.~\cite{Mei:2005gm}. There are, however, several caveats. Firstly, the fits are done to vertical-equivalent underground intensity data, which is known to only approximately correspond to the true vertical intensity at zenith angles below $\sim30^\circ$~\cite{Barrett:1952woo,Bugaev:1998bi,Fedynitch:2021ima}. In addition, various datasets are typically combined without compensating for systematic uncertainties through correction functions or nuisance parameters, and hence the fit results might be significantly biased towards experiments that have underestimated their uncertainties. They are also limited in the sense that they cannot provide additional physical information about the muons underground, such as their energy and angular distributions, which must be calculated using separate parametrizations.

The computational scheme of \mute{} (v1.0.1) used for labs under flat overburdens is described in Ref.~\cite{Fedynitch:2021ima}. The central quantity in the computation scheme from which all other underground and underwater observables are calculated is the underground flux, $\Phi^u$. This is calculated as a convolution between the surface muon fluxes from \mceq{} or \daemonflux{}, $\Phi^s$, and a surface-to-underground transfer tensor from \proposal{}, $U$, and is given by

\begin{equation}
\label{eq:convolution}
\Phi^u(E^u, X, \theta)=\Phi^s(E^s, \theta)*U(E^s, E^u, X),
\end{equation}

\noindent where $E^u$ is the underground muon energy, $X$ is the slant depth, $\theta$ is the zenith angle, $E^s$ is the surface muon energy, and $*$ denotes a convolution over the surface muon energies. The slant depth (or grammage) $X$ is the primary measure of distance in our calculations. Generally, it is defined as
\begin{equation}
    X(\ell, \theta, \phi) = \int_0^\ell\rho(\ell'(\theta, \phi))\,{\rm d}\ell', 
\end{equation}
where the integration is performed along the line of sight $\ell'$ from the center of the detector or lab, situated at the distance $\ell$ from the intersection with the surface in a specific direction, which is defined by $\theta$ and the azimuthal angle $\phi$. For homogeneous rock or for a given average density, this expression simplifies to
\begin{equation}
    \label{eq:slant_depth}
    X = \rho \ell \underset{\mathrm{flat}}{=} \rho\frac{d}{\cos{\theta}}.
\end{equation}
The second equivalence is valid for flat overburdens located at a fixed vertical depth $d$ in km. For convenience, it is common to convert distances and slant depths into kilometer water equivalent units (km.w.e.~$=10^5$~g\,cm$^{-2}$):

\begin{equation}
\label{eq:convert_to_kmwe}
\begin{aligned}
     X_{\rm km.w.e.}&=X / \rho_{\rm water} \\
     h_{\rm km.w.e.}&=d \rho/\rho_{\rm water}.
\end{aligned}
\end{equation}

The muon flux at the surface is assumed to have azimuthal symmetry. Furthermore, the altitude dependence of the surface flux can be safely neglected in our calculations since muons with energies relevant for underground fluxes at $h\geq 0.5$~km.w.e.\ (i.e.\ energies above $\sim100$~GeV) are produced in the upper atmosphere, around 10~km above sea level. For these reasons, we consider $\Phi^s$ to be a function only of $E^s$ and $\theta$. The conventions for definitions and notations used for ``flux" and ``intensity" in our work are explained in \cref{app:variabledefinitions}.

\begin{figure}
    \centering
    \includegraphics[width=\columnwidth]{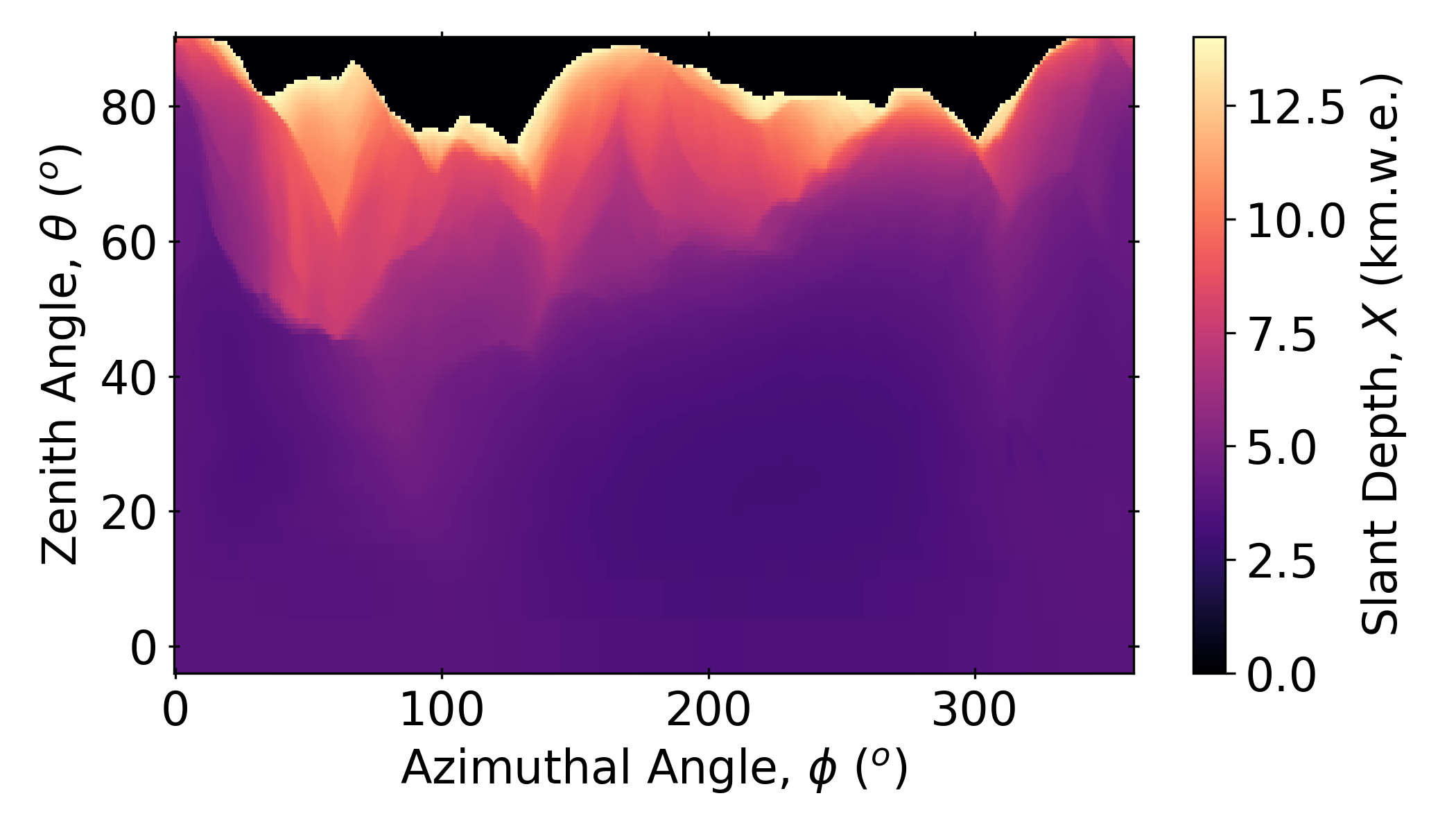}
    \caption{Slant depths of the Gran Sasso mountain, $X(\theta, \phi)$, in terms of the zenith and azimuthal angles~\cite{LVD:1998lir}. Depths greater than 14~km.w.e.\ have been masked (shown in black), as 14~km.w.e.\ is the default maximum slant depth for calculations by \mute{}. In all cases, muons traveling along $\theta=0^{\circ}$ are defined as vertical down-going muons.}
    \label{fig:gransassomountain}
\end{figure}

In this followup study, we introduce \mute{} v2~\cite{william_woodley_2022_6841971} and v3, capable of calculating muon observables for laboratories beneath mountains. For the case of a flat geometry above the lab (labs under flat overburdens or underwater), due to the assumed symmetry in the azimuthal angle, \cref{eq:slant_depth} provides a direct geometric correspondence between the slant depth and zenith angle for a given vertical depth. This relationship reduces the underground or underwater flux in \cref{eq:convolution} at the specified depth to a function of just two variables: $\Phi^u(E^u, \theta)$. For mountains, however, the nonuniform shape of the overburden introduces dependence on the azimuthal angle to the amount of rock a muon has to travel through, meaning the geometry of the mountain has to be taken into account in the calculations. To do this, topographic maps of the mountains in terms of the depth in spherical coordinates, $X(\theta, \phi)$, are obtained from the labs, such as the one for Laboratori Nazionali del Gran Sasso (LNGS) in Italy, shown in \cref{fig:gransassomountain}. For this reason, the slant depth and the zenith angle are kept as separate variables in the calculation of underground fluxes for mountains: $\Phi^u(E^u, X(\theta, \phi), \theta)$. The explicit dependence on $\theta$ is relevant because the surface fluxes depend on the zenith angle but not the azimuthal angle due to the symmetry at high energies.

For both overburden types (flat and mountainous), the differential underground muon intensity, $I^u$, is given by integrating the underground flux over the underground energy:

\begin{equation}
\label{eq:mountainintensity}
I^u(\theta, \phi)=\int_{E_{\mathrm{th}}}^{\infty}\Phi^u(E^u, X(\theta, \phi), \theta)\,\mathrm{d}E^u,
\end{equation}

\noindent where $E_{\mathrm{th}}$ is a choice for a threshold energy, typically defined by a specific detector. Since the typical energies of muons surviving the transit to deep underground labs end up in the GeV range underground, we set $E_{\mathrm{th}}$ to zero by default for all labs for generality. For a flat overburden lab situated at a fixed vertical depth, \cref{eq:mountainintensity} reduces to a function of one variable, $I^u(\theta)$. It remains a double-differential quantity for mountains as in \cref{eq:mountainintensity}, with $X(\theta, \phi)$ given by the topographical map.

We implement the calculation of the double-differential intensity in \mute{} v2 by precomputing an underground intensity matrix $I^u(X, \theta)$ on a constant grid of depths $X$ and zenith angles $\theta$. We then interpolate this distribution to a specific overburden profile $X(\theta, \phi)$ to obtain a matrix of lab-specific intensities $I^u(\theta, \phi)$. The elements of this matrix correspond to the bins of the mountain maps $X(\theta, \phi)$ provided directly by the experiments.

Using these underground intensities, we calculate various physical observables for different underground sites worldwide. The underground sites used are listed in \cref{tab:undergroundlabs}, along with their average rock density and vertical depths in km as found in the literature. We present flat overburden calculations for the Waste Isolation Pilot Plant (WIPP), Soudan Underground Laboratory, Boulby Underground Laboratory, Stawell Underground Physics Laboratory (SUPL), Sanford Underground Research Facility (SURF) located in the former Homestake Gold Mine, and the Sudbury Neutrino Observatory Laboratory (SNOLAB). For laboratories under mountains, we have used topographic maps of the Yangyang Underground Laboratory (Y2L) centered around the COSINE-100 detector, the Kamioka Observatory under the Ikenoyama mountain centered around two locations---the KamLAND~\cite{KamLAND:2009zwo} and Super-Kamiokande~\cite{KamLAND:2009zwo,Super-Kamiokande:2002weg} detectors---LNGS under the Gran Sasso mountain centered around the Large Volume Detector (LVD)~\cite{LVD:1998lir}, Laboratoire Souterrain de Modane (LSM) centered around the Fréjus detector~\cite{FREJUS:1989lko}, and China Jinping Underground Laboratory (CJPL-I) centered around the Jinping Neutrino Experiment (JNE)~\cite{JNE:2020bwn}. In the case of labs under mountains, it is necessary to center the map around a specific detector's location due to the calculations' dependence on the azimuthal angle. We selected these labs based on the availability of experimental data with which we can compare our results.

\begin{table}[]
\caption{\label{tab:undergroundlabs}Summary of underground sites used, their average rock density, $\rho$, their vertical depth in km, $d$, and their vertical depth in km.w.e., $h$. The depth in km.w.e. is calculated as $h=\rho d$. The sites are sorted by increasing equivalent vertical depth.}
\begin{ruledtabular}
\begin{tabular}{@{}lllr@{}}
Laboratory & Density, $\rho$ (g\,cm$^{-3}$) & $d$ (km) & $h$ (km.w.e.)\\
\colrule
WIPP & $2.3\pm 0.2$~\cite{Esch:2004zj} & 0.655~\cite{Esch:2004zj} & 1.507\\
Y2L & 2.7~\cite{Prihtiadi:2017sxc} & \multicolumn{2}{c}{Mountain}\\
Soudan & 2.85~\cite{ruddick1996underground} & 0.713~\cite{ruddick1996underground} & 2.032\\
Kamioka & $2.70\pm 0.05$~\cite{KamLAND:2009zwo,Hyper-Kamiokande:2018ofw} & \multicolumn{2}{c}{Mountain~\cite{KamLAND:2009zwo,Super-Kamiokande:2002weg}}\\
Boulby & $2.62\pm 0.03$~\cite{Robinson:2003zj} & 1.070~\cite{Robinson:2003zj} & 2.803\\
SUPL & 2.86~\cite{Urquijo:2016dxd} & 1.024~\cite{Barberio:2022grv} & 2.929\\
LNGS & $2.72\pm 0.05$~\cite{1998JAG....39...25C} & \multicolumn{2}{c}{Mountain~\cite{LVD:1998lir}}\\
LSM & $2.73\pm 0.01$~\cite{FREJUS:1989lko} & \multicolumn{2}{c}{Mountain~\cite{FREJUS:1989lko}}\\
SURF & $2.86\pm 0.11$~\cite{MAJORANA:2016ifg,Heise:2017rpu} & 1.478~\cite{MAJORANA:2016ifg,Heise:2017rpu} & 4.227\\
SNOLAB & $2.83\pm 0.05$~\cite{SNO:2009oor} & 2.092~\cite{SNO:2009oor} & 5.920\\
CJPL-I & 2.8~\cite{Wu:2013cno} & \multicolumn{2}{c}{Mountain~\cite{JNE:2020bwn}}\\
\end{tabular}
\end{ruledtabular}
\end{table}

By default, \mute{} provides surface-to-underground transfer tensors for muon energy loss in standard rock ($Z=11$, $A=22$, $\rho=2.65$~g\,cm$^{-3}$~\cite{menon1967progress,ParticleDataGroup:2022pth}), fresh water ($\rho=1.000$~g\,cm$^{-3}$), and sea water ($\rho=1.03975$~g\,cm$^{-3}$~\cite{koehne2013proposal}) for a range of slant depths spanning 0.5~km.w.e.\ to 14~km.w.e. The latest release, \mute{} v3, provides explicit transfer tensors for the specific rock density and composition at the locations of the labs listed in~\cref{tab:undergroundlabs}, resulting in a more precise and reliable muon flux calculation.

\section{\label{sec:rock_composition}Modeling Rock Composition}
The flux of muons underground and underwater is governed largely by the energy losses of the muons as they travel through rock and water. When traveling through matter, high-energy muons relevant to underground experiments lose energy via pair production, bremsstrahlung, photonuclear interactions, and ionization. The cross sections for these interactions depend on the medium's chemical composition, making accurate modeling of media a significant factor in flux calculations. Media in \proposal{} are modeled by three sets of values:

\begin{enumerate}[label=\Alph*.]
    \item The rock density,
    \item The rock composition,
    \item The Sternheimer parameters.
\end{enumerate}

\noindent The effects of each of these and how they are implemented into our calculations with \mute{} v3 are briefly explained in the following subsections.

\subsection{Rock density}
All internal \proposal{} calculations for energy loss are performed in units of grammage, meaning they are agnostic to changes in the density. The rock density, therefore, only enters the calculations when converting physical depths in km to grammage in km.w.e.\ via \cref{eq:convert_to_kmwe}.

For the labs under flat earth in \cref{tab:undergroundlabs}, the rock density has been accounted for in the lab's vertical depth, $h$. This has sometimes led to different vertical depths being used in this work compared to those found in the literature. For the labs under mountains, some mountain maps were already available in units of km.w.e., in which case no conversions were made. However, when the slant depths in the mountain profile maps were given in units of km, they were converted into km.w.e.\ of laboratory rock by multiplying by the corresponding rock density in \cref{tab:undergroundlabs}. This simple linear scaling of the slant depths is an estimate of the impact of the average rock density alone, as chemical differences between different rock types are considered only for energy losses and not when calculating depths. This is in contrast to parametric polynomials given by some experiments, such as LVD~\cite{LVD:1998lir} and SNO~\cite{SNO:2009oor}, for example. These experiments use underground muon intensity data to construct conversion formulas from standard rock to laboratory rock, which aim to account for all differences between the rock types when converting depths.

\begin{figure}
    \centering
    \includegraphics[width=\columnwidth]{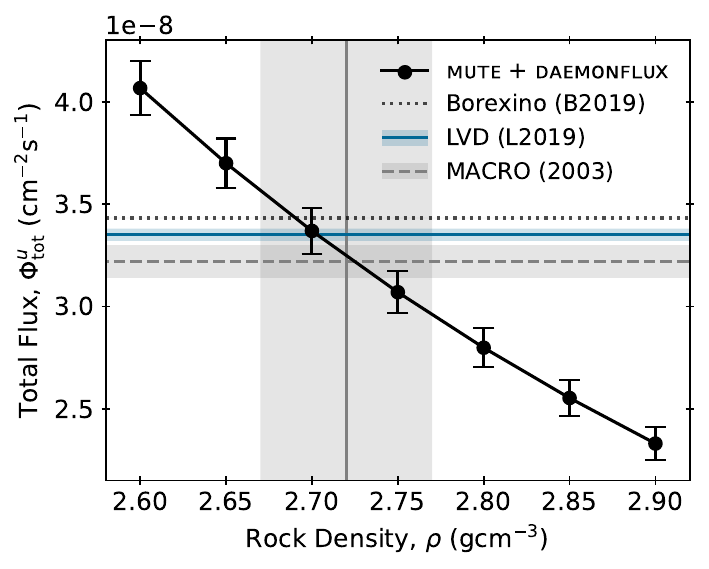}
    \caption{Total underground muon fluxes as a function of rock density under the Gran Sasso mountain centered around LVD in Hall A of LNGS. The error bars represent the uncertainty from \daemonflux{}. The horizontal lines and their error bands represent the measurements from MACRO (Hall B)~\cite{MACRO:2002qsl}, Borexino (Hall C)~\cite{Borexino:2018pev}, and LVD (Hall A)~\cite{LVD:2019zlh}. The vertical line and its error band represent the measured rock density of $(2.72\pm 0.05)$~g\,cm$^{-3}$, given in \cref{tab:undergroundlabs}.}
    \label{fig:flux_vs_density}
\end{figure}

Although rock density is essential for calculating total muon fluxes underground, many labs have significant uncertainties on their rock density, or no uncertainties at all, as seen in \cref{tab:undergroundlabs}. Additionally, a laboratory can misjudge its rock density if too little information on the rock composition is available. Because of the relation between rock density and underground muon flux and the precision of the results produced by \mute{}, there is potential for \mute{} to constrain uncertainties on rock densities or even correct misjudged rock densities. We have done this for LNGS in \cref{fig:flux_vs_density}, and compare the results to three experimental measurements from MACRO, Borexino, and---the most recent LNGS measurement---LVD. The good agreement with LVD suggests that it is likely possible to further constrain rock densities for other labs using \mute{}; however, we note that the precision of the surface flux model is not yet good enough to resolve the slight differences between different experimental halls.

\subsection{Rock composition}
The cross sections for muon energy loss interactions are dependent on the average number of protons, $\left<Z\right>$, and the average atomic mass, $\left<A\right>$, of the propagation medium. The cross sections are proportional to $\left<Z^2/A\right>$ for pair production~\cite{1971ICRC....6.2436K} and bremsstrahlung~\cite{Kelner:1995hu,Kelner:1999yf}, to $\left<A\right>$ for photonuclear interactions~\cite{Abramowicz:1991xz,Abramowicz:1997ms}, and to $\left<Z\right>$ for ionization~\cite{Rossi:1952kt}. Although overburdens are typically made up of multiple types of rock, for the optimization of the program, homogeneous media defined by average $\left<Z/A\right>$ and $\left<Z^2/A\right>$ values have been used in \proposal{}. The definitions of these average values are given in \cref{app:rockcompositions}, and the values used for each lab are listed in \cref{tab:rock_compositions}.

\begin{table*}[]
\caption{\label{tab:rock_compositions}Rock composition definitions in terms of average numbers of protons and nucleons. In some cases, multiple contrasting values for $\left<Z\right>$ and $\left<A\right>$ are found in or derived from the literature, as is the case with Soudan (see Refs.~\cite{Mei:2005gm,MINOS:2007laz,ruddick1996underground}), Kamioka (see Refs.~\cite{KamLAND:2009zwo,Mizukoshi:2018rnt}), LSM (see Refs.~\cite{FREJUS:1989lko,Chazal:1997qn,rhode1993study}), and SNOLAB (see Refs.~\cite{kyba2006measurement,Ewan:1987oqj,SNO:2009oor}). In these cases, the set of values for which the most detailed information was provided was used.}
\begin{ruledtabular}
\begin{tabular}{@{}llllll@{}}
Laboratory & $\left<Z\right>$ & $\left<A\right>$ & $\left<Z/A\right>$ & $\left<Z^2/A\right>$ & Reference\\
\colrule
Standard Rock & 11.00 & 22.00 & 0.50 & 5.50 & \cite{ParticleDataGroup:2022pth}\\
WIPP\footnote{Because WIPP is a salt mine, its overburden consists primarily of NaCl. Therefore, the full definition of salt, including its Sternheimer parameters in \cref{tab:sternheimer} in \cref{app:rockcompositions}, was used for WIPP. The $\left<Z\right>$ and $\left<A\right>$ values listed here are slightly different from those found in Ref.~\cite{osti_6441454} (where $\left<Z\right>=14.64$ and $\left<A\right>=30.00$) due to the presence of other elements in the rock samples in Ref.~\cite{osti_6441454}.} & 14.00 & 29.25 & 0.49 & 7.14 & \cite{koehne2013proposal}\\
Y2L & 11.79 & 23.79 & 0.50 & 5.85 & \cite{Yoon:2021tkv}\\
Soudan & 12.32 & 24.90 & 0.50 & 6.10 & \cite{ruddick1996underground}\\
Kamioka & 11.31 & 22.76 & 0.50 & 5.62 & \cite{Mizukoshi:2018rnt}\\
Boulby & $11.70\pm 0.50$ & 
$23.60\pm 1.00$ & $0.50\pm 0.03$ & $5.80\pm 0.55$ & \cite{Robinson:2003zj}\\
LNGS & 11.42 & 22.83 & 0.50 & 5.71 & \cite{MACRO:1990ykz,MACRO:1995egd}\\
LSM\footnote{The full definition of Fréjus rock, including Sternheimer parameters from Ref.~\cite{koehne2013proposal} in \cref{tab:sternheimer}, was used for LSM.} & 11.74 & 23.48 & 0.50 & $5.87\pm 0.02$ & \cite{FREJUS:1989lko}\\
SURF & 12.01 & 23.98 & 0.50 & 6.01 & \cite{Mei:2009py}\\
SNOLAB & 12.02 & 24.22 & 0.50 & 5.96 & \cite{Ewan:1987oqj}\\
CJPL-I & 12.15 & 24.30 & 0.50 & 6.07 & \cite{CDEX:2021cll}\\
\end{tabular}
\end{ruledtabular}
\end{table*}

$\left<Z/A\right>$ and $\left<Z^2/A\right>$ values are sometimes found directly in the literature, though, in some cases, only percent weights for minor component chemical compositions are found. In these latter cases, the average numbers of protons and nucleons were calculated from these compositions. All available chemical compositions are listed in \cref{tab:geochemical} in \cref{app:rockcompositions}. Lastly, sometimes values are published as $\left<Z\right>$ and $\left<A\right>$ instead of $\left<Z/A\right>$ and $\left<Z^2/A\right>$. In general, $\left<Z/A\right>\neq\left<Z\right>/\left<A\right>$, meaning algebraic conversion from one pair of values to the other is not possible; however, the approximation error for this inequality is sufficiently small (less than 1\% in all cases) that it is neglected, and we approximate $\left<Z/A\right>=\left<Z\right>/\left<A\right>$ where needed to fill \cref{tab:rock_compositions}.

A detailed geological survey of the Stawell, Victoria area in Australia was published in Ref.~\cite{DUGDALE201041}, but no information on the average rock composition is provided. It is known that the rock above SUPL is mainly basalt~\cite{Urquijo:2016dxd,Barberio:2022grv,Zurowski:2023wwx}, but the chemical composition of different types of basalt can vary widely, with the percent weight of SiO$_2$ lying between 45\% and 52\%, which can lead to significant variations in the energy loss and therefore total underground fluxes. Due to a lack of precise geochemical measurements, this work used standard rock for all SUPL calculations.

\begin{figure}
    \centering
    \includegraphics[width=\columnwidth]{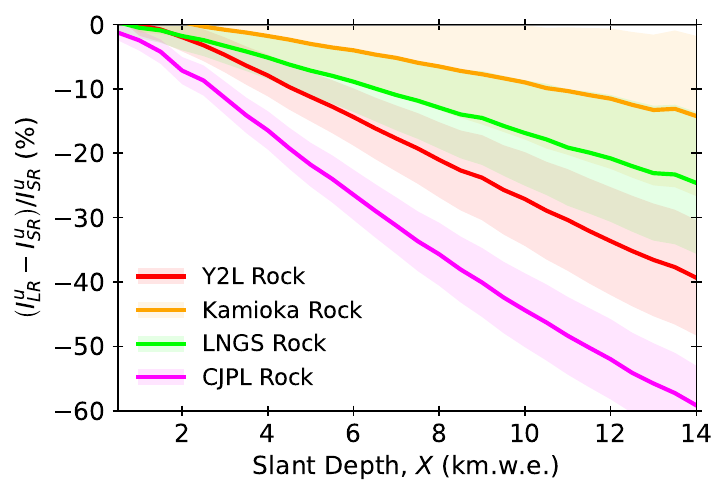}
    \includegraphics[width=\columnwidth]{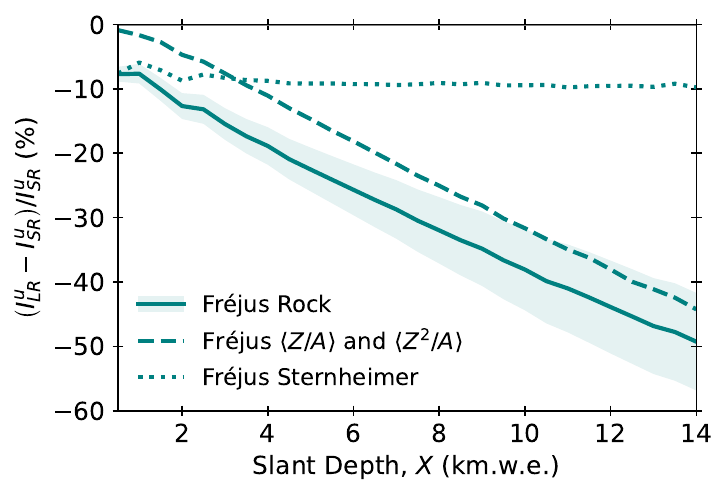}
    \caption{Deviation of underground muon intensities calculated using laboratory rock (LR) types from \cref{tab:rock_compositions} from intensities calculated using standard rock (SR) vs slant depth for four labs under mountains (top) and LSM (bottom). In both panels, intensities calculated using standard rock $\left<Z/A\right>$ and $\left<Z^2/A\right>$ and Sternheimer parameters are represented by $I^u_{\mathrm{SR}}$. The top panel shows the effects of changing solely the $\left<Z/A\right>$ and $\left<Z^2/A\right>$ values in the \proposal{} simulations, whereas the bottom panel shows the effect of changing both the $\left<Z/A\right>$ and $\left<Z^2/A\right>$ values and the Sternheimer parameters. Error bands in both panels represent the uncertainty from the \daemonflux{} model. Lab depths and mountain profiles are not taken into account.}    
    \label{fig:rock_composition}
\end{figure}

The effect of changing the $\left<Z/A\right>$ and $\left<Z^2/A\right>$ values on the energy loss results from \proposal{} can be significant. \Cref{fig:rock_composition} shows the ratio of underground muon intensities using the laboratory rock from \cref{tab:rock_compositions} to those using standard rock, demonstrating that increasing the $\left<Z^2/A\right>$ value leads to a decrease in the muon intensity results, as physically expected. There is a deviation from standard rock of -27\% in the underground intensity for CJPL-I at the vertical depth of the lab (approximately 6~km.w.e.), while the effect is minimal for Y2L, Kamioka, and LNGS at their respective vertical depths (less than -5\%). However, the effect is amplified as slant depth increases, as the muons travel through higher quantities of rock. Therefore, muons reaching the labs at higher zenith angles are affected more greatly by changes to the chemical composition. Because of this depth-dependent energy loss effect, DIRs and mean underground energy parametrizations that are not corrected for rock type, like those in \cref{eq:DIR_PDG,eq:DIR_LVD}, are less suitable for high-precision calculations, highlighting the need for detailed simulations in underground muon flux calculations.

Lastly, it should be noted that because the values presented in \cref{tab:rock_compositions} are averages, and overburdens typically contain various types of rocks in different quantities, the $\left<Z/A\right>$ and $\left<Z^2/A\right>$ values will contain some margin of error that is not taken into account. While these average values are assumed to be sufficient approximations for the definitions of the propagation media, more detailed information about rock compositions can be helpful for more precise results.

\subsection{Sternheimer parameters}
To describe ionization losses, \proposal{} uses the Bethe--Bloch equation~\cite{ParticleDataGroup:2022pth}. This equation includes a density correction term, $-\delta/2$, as described in Refs.~\cite{Sternheimer:1952jn,koehne2013proposal}, where $\delta$ is given by

\begin{equation}
\label{eq:sternheimerparameters}
\delta=\begin{cases}
\displaystyle
\delta_0 10^{2(x-x_0)} & \text{if $x<x_0$,}\\
2\ln(10)x+c+a(x_1-x)^b &\text{if $x_0\leq x\leq x_1$,}\\
2\ln(10)x+c &\text{if $x>x_1$.}
\end{cases}
\end{equation}

\noindent Here, $x_0$ and $x_1$ demarcate transition regions of the function form of the $\delta$ parameter~\cite{Sternheimer:1952jn}. $\delta_0$, $a$, $b$, and $c$ are additional empirical parameters, all together termed the ``Sternheimer parameters." Values for these parameters are typically taken from~\cite{Sternheimer:1983mb,lohmann1985energy,Groom:2001kq,koehne2013proposal}, and \cref{tab:sternheimer} in \cref{app:rockcompositions} lists the values for the media used in this study.

Although important for energy loss, the only relevant rock type for which published Sternheimer parameters exist in the literature (in addition to standard rock) is Fréjus rock, the rock above LSM~\cite{koehne2013proposal}. For this reason, the calculations for LSM presented in this work were done using the full definition of Fréjus rock. For other laboratories, however, because values for the Sternheimer parameters do not exist in the literature, all calculations with \mute{} use the Sternheimer parameters for standard rock as an approximation. This is with the exception of WIPP, which uses an overburden of salt.

The difference in energy loss between standard rock and Fréjus rock is shown in the bottom panel of \cref{fig:rock_composition}. This panel shows the effect of using the Fréjus $\left<Z/A\right>$ and $\left<Z^2/A\right>$ values with the dashed line, the impact of using the Fréjus Sternheimer parameters with the dotted line, and the effect of using both with the solid line. The Sternheimer parameters are shown to contribute a nearly constant 10\% decrease in underground muon intensity across all depths. Therefore, we estimate at least a 10\% systematic error on predictions for other laboratories due to the lack of knowledge of Sternheimer parameters.

\section{Total Muon Flux}
\label{sec:total_underground_muon_flux}
The total muon flux is the main physical observable of interest for underground and underwater experiments for the purposes of muon-induced background studies. It is calculated by integrating the underground or underwater intensity from \cref{eq:mountainintensity} over the solid angle:

\begin{equation}
\label{eq:total_flux}
\Phi^u_{\mathrm{tot}}=\iint_{\Omega}I^u(X(\theta, \phi), \theta)\,\mathrm{d}\Omega.
\end{equation}

\noindent We have performed this calculation for labs underground and underwater and present the results below.

\subsection{Total flux underground}
\begin{figure}
    \centering
    \includegraphics[width=\columnwidth]{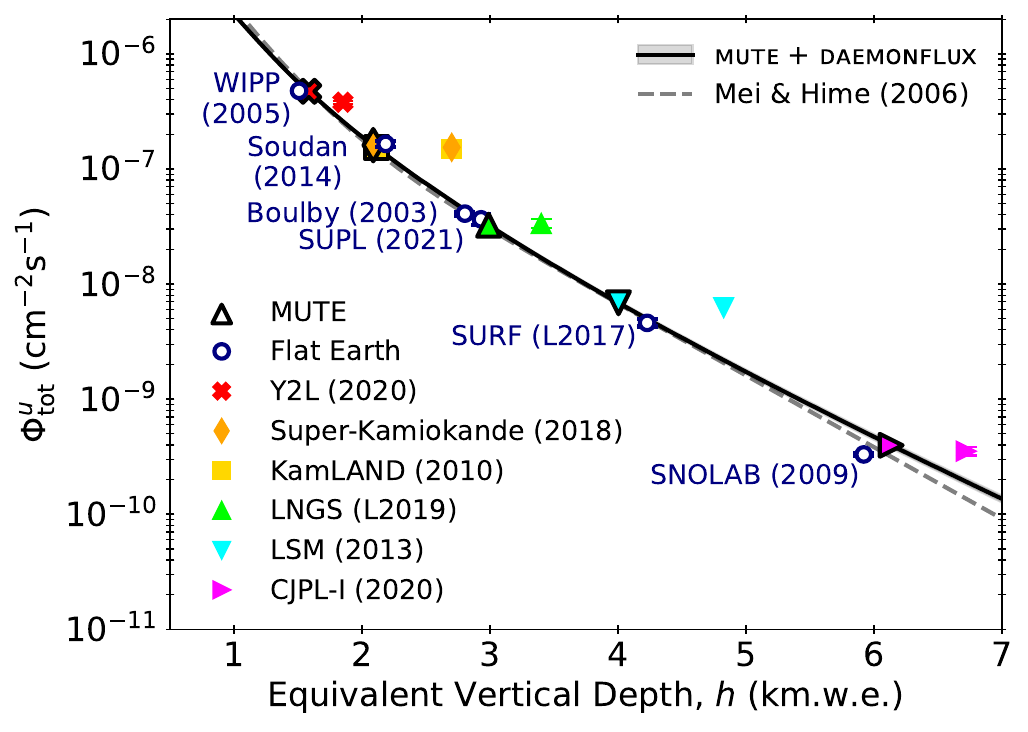}
    \caption{Total underground muon flux vs equivalent vertical depth underground. The \mute{} curve was calculated for a flat overburden of standard rock, and the points with thick black outlines were calculated with \mute{} using topographic maps and rock compositions of the mountain. The latter have had their depth values fitted to lie on the \mute{} curve, giving the laboratory's equivalent vertical depth. Data points are shown for the most recent experimental flux values listed in \cref{tab:depthsandtotalfluxes} at the vertical depths of the labs quoted in the literature. These depths are sometimes given as minimum, average, or straight vertical depths (``engineering depths"), rather than the equivalent vertical depths used in this work. The curve from the parametric formula in \cref{eq:DIR_PDG} from Ref.~\cite{Mei:2005gm} is also shown for comparison.}    
    \label{fig:total_flux}
\end{figure}

The total underground flux has been calculated for the sites listed in \cref{tab:undergroundlabs} using \daemonflux{} as the surface muon flux model and U.S.\ Standard Atmosphere~\cite{atmosphere1976national} as the atmospheric density model. The results are given in \cref{tab:depthsandtotalfluxes} and are shown in \cref{fig:total_flux} along with experimental measurements.

\begin{table*}[]
\caption{\label{tab:depthsandtotalfluxes}Experimental and predicted total muon fluxes for underground laboratories. The predicted total flux values were used to infer equivalent vertical depths by allowing the \mute{} calculation to float along the $x$ axis and fitting it to the curve for standard rock in \cref{fig:total_flux}, resulting in a standard-rock equivalent vertical depth, $\overline{h}_{\mathrm{SR}}$, for all labs.}
\begin{ruledtabular}
\begin{tabular}{@{}lllll@{}}
\multirow{2}{*}{Laboratory} & \multirow{2}{*}{Experiment} & \multicolumn{1}{c}{$\Phi^u_{\mathrm{tot}}$ (cm$^{-2}$\,s$^{-1}$)} & \multicolumn{1}{c}{$\Phi^u_{\mathrm{tot}}$ (cm$^{-2}$\,s$^{-1}$)} & \multicolumn{1}{c}{$\overline{h}_{\mathrm{SR}}$ (km.w.e.)}\\
& & \multicolumn{1}{c}{Measured} & \multicolumn{1}{c}{Predicted by \mute{}} & \multicolumn{1}{c}{Inferred from \mute{}}\\
\colrule
WIPP & - (2005) & $(4.77\pm 0.09)\times 10^{-7}$~\cite{Esch:2004zj} & $(5.17\pm 0.11)\times 10^{-7}$~\footnote{Calculated using Sternheimer parameters other than those for standard rock (see \cref{tab:sternheimer,tab:rock_compositions}).} & $1.54\pm 0.01$\\

\multirow{2}{*}{Y2L} & \multirow{2}{*}{COSINE-100 (2020)} & $(3.795\pm 0.110)\times 10^{-7}$~\cite{COSINE-100:2020jml} & \multirow{2}{*}{$(4.73\pm 0.11)\times 10^{-7}$} & \multirow{2}{*}{$1.58\pm 0.01$}\\

& & $(4.459\pm 0.132)\times 10^{-7}$~\footnote{Corrected for multiple muons (see text).} & & \\

Soudan & - (2014) & $(1.65\pm 0.10)\times 10^{-7}$~\cite{Zhang:2014jsq} & $(1.66\pm 0.04)\times 10^{-7}$ & $2.07\pm 0.01$\\

Kamioka & Super-Kamiokande (2018) & $(1.54\pm 0.31)\times 10^{-7}$~\cite{Hyper-Kamiokande:2018ofw}~\footnote{Calculated from simulation, not from experimental data.} & $(1.61\pm 0.04)\times 10^{-7}$ & $2.09\pm 0.01$\\

& KamLAND (2010) & $(1.49\pm 0.11)\times 10^{-7}$~\cite{KamLAND:2009zwo} & $(1.53\pm 0.04)\times 10^{-7}$ & $2.11\pm 0.01$\\

Boulby & ZePLiN 1 (2003) & $(4.09\pm 0.15)\times 10^{-8}$~\cite{Robinson:2003zj} & $(4.19\pm 0.13)\times 10^{-8}$ & $2.83\pm 0.02$\\

SUPL & SABRE (2021) & $(3.65\pm 0.41)\times 10^{-8}$~\cite{Melbourne:2021wxo} & $(3.58\pm 0.11)\times 10^{-8}$~\footnote{Calculated using standard rock rather than being tuned to the rock above the laboratory.} & $2.93\pm 0.02$\\

LNGS & MACRO (2003) & $(3.22\pm 0.08)\times 10^{-8}$~\cite{MACRO:2002qsl} & \multirow{3}{*}{$(3.25\pm 0.11)\times 10^{-8}$} & \multirow{3}{*}{$2.99\pm 0.02$}\\

& Borexino (B2019) & $(3.432\pm 0.003)\times 10^{-8}$~\cite{Borexino:2018pev} & & \\

& LVD (L2019) & $(3.35\pm 0.03)\times 10^{-8}$~\cite{LVD:2019zlh} & & \\

LSM & EDELWEISS (2013) & $(6.25\pm 0.2^{+0.6}_{-1.0})\times 10^{-9}$~\cite{EDELWEISS:2013kzp} & $(6.87\pm 0.28)\times 10^{-8}$~\footnotemark[1] & $4.00\pm 0.03$ \\

SURF & Homestake (1983) & $(4.14\pm 0.05)\times 10^{-9}$~\cite{Cherry:1983dp} & \multirow{3}{*}{$(4.01\pm 0.17)\times 10^{-9}$} & \multirow{3}{*}{$4.38\pm 0.03$}\\

& MAJORANA (M2017) & $(5.31\pm 0.17)\times 10^{-9}$~\cite{MAJORANA:2016ifg} & & \\

& LUX (L2017) & $(4.60\pm 0.33)\times 10^{-9}$~\cite{ihm2018through} & & \\

SNOLAB & SNO (2009) & $(3.31\pm 0.10)\times 10^{-10}$~\cite{SNO:2009oor} & $(4.02\pm 0.24)\times 10^{-10}$ & $6.13\pm 0.05$\\

CJPL-I & JNE (2020) & $(3.53\pm 0.29)\times 10^{-10}$~\cite{JNE:2020bwn} & $(3.98\pm 0.24)\times 10^{-10}$ & $6.13\pm 0.05$\\
\end{tabular}
\end{ruledtabular}
\end{table*}

In \cref{fig:total_flux}, the total underground flux is plotted against ``equivalent vertical depth." For labs under flat overburdens, this is the depth of the lab, comparable to depths listed in \cref{tab:undergroundlabs}. For labs under mountains, however, the equivalent vertical depth is defined, as in Ref.~\cite{Mei:2005gm}, as the depth a lab would be at under a flat overburden given the total underground muon flux seen by the lab. We allow the \mute{} points for labs under mountains in \cref{fig:total_flux} to float along the $x$ axis and fit their depths so they lie directly on the \mute{} curve, which has been calculated for standard rock. We report these inferred depths in \cref{tab:depthsandtotalfluxes} and label them $\overline{h}_{\mathrm{SR}}$. The uncertainties on these values come from the error band on the \mute{} curve.

\begin{figure}
    \centering
    \includegraphics[width=\columnwidth]{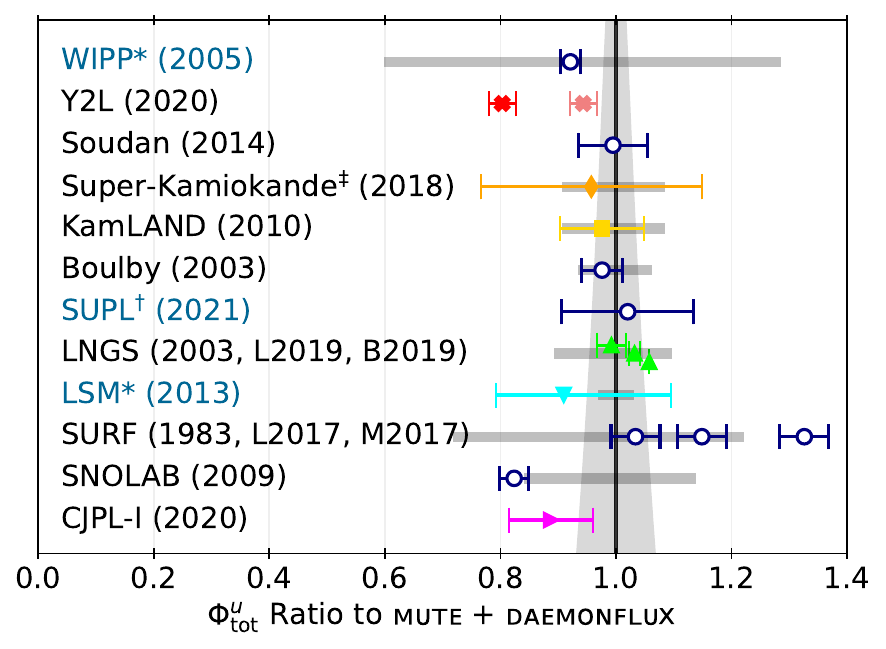}
    \caption{The ratio of experimental total underground muon flux measurements to \mute{}, using the results in \cref{tab:depthsandtotalfluxes}. The error band represents the uncertainty in the model from \daemonflux{}, while the horizontal bars represent uncertainty from the rock density values given in \cref{tab:undergroundlabs}. Data points are shown for the experimental flux values listed in \cref{tab:depthsandtotalfluxes}. An asterisk ($^*$) indicates the calculation uses Sternheimer parameters other than those for standard rock (see \cref{tab:sternheimer,tab:rock_compositions}). A dagger ($^{\dagger}$) indicates that our calculation used standard rock rather than being tuned to the rock above the given laboratory. A double dagger ($^{\ddag}$) indicates that the point used is a prediction calculated from simulation and does not come from experimental data. Where multiple data points are shown, the left-to-right order of the points is the same as the order of the measurements in brackets in the labels. Similar ratios produced with other surface flux models can be found in \cref{app:comparisons_to_additional_models}.}
    \label{fig:total_flux_ratio}
\end{figure}

The official data points shown in \cref{fig:total_flux} for labs under mountains are seen to be offset from the \mute{} result along the $x$ axis, such as the CJPL-I lab under the Jinping mountain, whose depth we infer to be $(6.20\pm 0.13)$~km.w.e. This is notably different from the value of 6.72~km.w.e.\ quoted in, for example, Refs.~\cite{JNE:2020bwn,Wan:2019qml}, which is stated to be the result of a scaling factor $F$ from cosmic ray leakage due to the topographic profile of the mountain. 6.72~km.w.e.\ is, in fact, the straight vertical depth above the lab (termed the ``engineering depth"), not the equivalent vertical depth as used in our work. This is also the reason for the offset between the experimental and calculated points for KamLAND and Super-Kamiokande, while a similar offset is present for LNGS and LSM caused by the use of average vertical depths in the literature. The inconsistencies in the literature in defining a meaningful vertical depth for laboratories under mountains make the results difficult to visually interpret. Therefore, \cref{fig:total_flux_ratio} shows the ratio of the results in \cref{fig:total_flux} to the \mute{} predictions in \cref{tab:depthsandtotalfluxes} independent of lab depth.

The main uncertainties in \mute{} come from the surface flux model, \daemonflux{}, and the modeling of the overburden. The former are represented by the gray error bands in \cref{fig:total_flux,fig:total_flux_ratio}, and contribute an error of $\sim$7\% for deep depths (higher muon energies) and around 2\% for shallow depths. The latter are represented by horizontal gray bars and are entirely attributed to the range of allowed average rock densities listed in \cref{tab:undergroundlabs}. These two errors are shown separately to emphasize their independence from each other. The interaction model uncertainties are independent of the lab, in contrast to the knowledge of the rock density, which is either limited by the precision of the geological surveys or by the nonuniformity of the overburden, making it difficult to compute a representative mean density. The interaction model errors have been discussed explicitly in Ref.~\cite{Fedynitch:2021ima}, where the rock density variations were implicitly included in the errors of the experimental data. The magnitudes of the uncertainties in \cref{tab:undergroundlabs} are large in many cases, leading to significant variations in the total flux seen in \cref{fig:total_flux_ratio}. Error bars are not shown for labs where no uncertainty on the rock density was found in the literature. The mountain maps cited in Sec. \ref{sec:computationalmethod} are assumed to be exact, so no error was considered for the slant depth values. Lastly, the \proposal{} Monte Carlo simulations in \mute{} were performed with $N=10^6$ events per energy and zenith angle bin, making the statistical error across all calculations negligible.

In general, discrepancies between predictions and measurements are not unexpected because the measurement conditions are not always possible to fully reproduce. Many factors can have a significant effect on the results, including the composition of the rock, the location of the lab under the mountain, the location of the detector in the lab, the energy threshold of the detector, the angular acceptance of the detector, and analysis decisions like energy cuts, binning choices, whether the analysis is inclusive of muon bundles or just single muons, and the treatment of seasonal variations in the collection and analysis of the data. More detailed information and analyses of these factors for a given detector may further improve the accuracy of \mute{} results. Although these details were not always available for each laboratory for this study, \mute{} is provided as an open-source tool for experiments to tailor to their specific use cases.

Despite the many potential factors contributing to the accuracy of the calculations, \mute{} in combination with \daemonflux{} provides an accurate description of the underground muon flux data within uncertainties for nearly every lab. The precision of the calculations of the total underground flux and the equivalent vertical depth in \cref{tab:depthsandtotalfluxes} are good enough, for example, to resolve the difference in location between the KamLAND and Super-Kamiokande detectors under the Ikenoyama mountain. This agreement between prediction and measurement is also indicated when comparing our results to measurements of the muon seasonal variation amplitudes in one of our previous works~\cite{Woodley:2023gnf}. As a result, the total underground muon flux may be one of the most robust observables to use to constrain primary cosmic ray flux models or hadronic yields because the results are model-independent. However, because of the magnitudes of the systematic uncertainties coming from the rock density errors in \cref{fig:total_flux_ratio}, the constraints provided to flux model calibrations might not be as strong as they potentially could be~\cite{Yanez:2023lsy,Honda:2019ymh}.

In the next subsections, we discuss in more detail the results shown in \cref{fig:total_flux_ratio} for a subset of laboratories.

\subsubsection{Y2L}
The clearest outlier in \cref{fig:total_flux_ratio} is the measurement for Y2L. Multiple measurements have been published for the COSINE-100 detector in the A5 hall of Y2L~\cite{COSINE-100:2017dsl,Prihtiadi:2017sxc,COSINE-100:2020jml}, all of which are in close agreement with each other but in disagreement with the \mute{} prediction by about 20\%, according to \cref{fig:total_flux_ratio}. An additional measurement from the KIMS experiment has been published in Refs.~\cite{kims2005,Kims:2005dol}, but KIMS is located in the A6 hall of Y2L, which is displaced 200~m away from the A5 hall~\cite{COSINE-100:2017dsl}. As seen with the difference in \mute{} results for the KamLAND and Super-Kamiokande detectors, which are spaced 150~m apart in the Kamioka laboratory~\cite{Pronost:2018ghn}, \mute{} calculations are sensitive enough to capture these differences in the positions of detectors. Therefore, because the mountain map we received for Y2L is centered around the COSINE-100 detector in the A5 hall, we compare only to the most recent COSINE-100 measurement from Ref.~\cite{COSINE-100:2020jml}.

A possible explanation for the discrepancy between our prediction and the COSINE-100 result is muon bundles. The flux calculated by \mute{} corresponds to that of single muons; however, the COSINE-100 detector and analysis make no distinction between single and multiple muons. According to \cref{tab:depthsandtotalfluxes}, the equivalent vertical depth of Y2L is 1.58~km.w.e. Using \mute{} in combination with \mceq{}, we approximate that, at this depth, 15\% of events are expected to contain two or more muons per event. An excess of events in Fig.\ 9 in Ref.~\cite{COSINE-100:2017dsl} suggests an approximate contribution of 10\% due to muon bundles captured by the detector within detector acceptance. This implies a total muon flux that would be 15--20\% higher than reported. We have computed a correction for this in \cref{tab:depthsandtotalfluxes} and have indicated this corrected value by a pale pink point for Y2L in \cref{fig:total_flux_ratio}.

\subsubsection{Soudan}

Our result of $(1.66\pm 0.04)\times 10^{-7}$\,cm$^{-2}$s$^{-1}$ for the total underground flux in the Soundan Underground Laboratory is in very good agreement with the measurement of $(1.65\pm 0.10)\times 10^{-7}$\,cm$^{-2}$s$^{-1}$ from Ref.~\cite{Zhang:2014jsq}. We note that Ref.~\cite{Zhang:2014jsq} is the only published peer-reviewed measurement for Soudan, though we find good to moderate agreement with other Soudan measurements as well, including a 1997 Soudan 2 value of $(1.80\pm 0.09)\times 10^{-7}$\,cm$^{-2}$s$^{-1}$~\cite{ruddick1996underground}, an often-quoted folded value of $(2.0\pm 0.2)\times 10^{-7}$\,cm$^{-2}$s$^{-1}$~\cite{Kamat:2005ct}, and an unpublished MINOS measurement of $1.77\times 10^{-7}$\,cm$^{-2}$s$^{-1}$~\cite{Zhang:2014jsq}.

It should be noted that the agreement we find is highly dependent on the rock density and vertical depth used for the lab. There is, however, wide variation in the depths of the lab quoted in the literature; see Refs.~\cite{ruddick1996underground,MINOS:2007laz,MINOS:2015ikk,Zhang:2014jsq,CDMS:2005jsf,MINOS:2009njg}, all of which give different vertical depths for the same level of the laboratory, ranging from 0.690~km to 0.780~km, and Refs.~\cite{MINOS:2007laz,MINOS:2020iqj,MINOS:2009njg,CDMS:2005jsf,MINOS:2015ikk,Zhang:2014jsq,Soudan-2:1996hoz}, which give vertical depths ranging from 2.07~km.w.e.\ to 2.10~km.w.e. These differences in depth lead to a maximum possible discrepancy of over 20\% between experimental measurements and our \mute{} prediction in some cases. Therefore, for consistency, in \cref{tab:undergroundlabs} and in our calculations, we used the Soudan 2 density and vertical depth in km given in Ref.~\cite{ruddick1996underground} to calculate a vertical depth in km.w.e. We then used that vertical depth in km.w.e.\ to calculate the total flux for a flat overburden and we compare that result to the measurement from Ref.~\cite{Zhang:2014jsq}.

Furthermore, as both Ref.~\cite{Zhang:2014jsq} and Ref.~\cite{ruddick1996underground} note, muon intensity changes as location in the laboratory, altitude, and rock density change, and there is significant variation in the rock density and composition above Soudan. For this reason, more accurate information on the overburden is needed for a detailed comparison. Although Fig.\ 10 of Ref.~\cite{Zhang:2014jsq} shows a variation in the mountain profile above the lab of around 0.1~km, which is enough to have a great effect on \mute{} results, we performed the calculation for a flat overburden because we had no access to this topographic map. Despite this, we still see overall very good agreement with the Ref.~\cite{Zhang:2014jsq} measurement.

\subsubsection{SUPL}
As stated in Sec. \ref{sec:rock_composition}, aside from the rock density, no additional relevant quantitative information exists about the rock above SUPL. For this reason, standard rock was used in the \proposal{} simulations. Despite not considering the rock composition, the \mute{} result agrees well with the SABRE measurement from Ref.~\cite{Melbourne:2021wxo}. We note that the result may change with more information about the rock density and composition.

\subsubsection{SURF}
Three total underground muon flux measurements have been published for SURF, from Homestake~\cite{Cherry:1983dp}, MAJORANA~\cite{MAJORANA:2016ifg}, and LUX~\cite{ihm2018through}, all taken at the 4850-level Davis laboratory location.

Detailed information on the rock's chemical composition above SURF is available in the literature~\cite{Heise:2017rpu,caddey1991homestake,hart2014topographic,Mei:2009py}. However, because of the significant variation in different rock types, with densities ranging from $2.43$~g\,cm$^{-3}$ to $3.26$~g\,cm$^{-3}$, the uncertainty on the rock density remains high. With the uncertainties from the \daemonflux{} model, the rock density, and the experimental systematics all taken into account, we find the best agreement with Homestake and the worst with MAJORANA.

It can be seen in \cref{fig:total_flux_ratio} that these three measurements are not in good agreement with each other. Possible reasons for these disagreements are discussed in Ref.~\cite{ihm2018through}. For the case of the Homestake measurement, Ref.~\cite{Cherry:1983dp} finds a single-muon intensity of $(4.91\pm 0.06)\times 10^{-9}$~cm$^{-2}$\,s$^{-1}$\,sr$^{-1}$ for zenith angles up to $18^{\circ}$, where the uncertainty is solely statistical. Reference~\cite{ihm2018through}, however, corrects this to a value of $(4.14\pm 0.05)\times 10^{-9}$~cm$^{-2}$\,s$^{-1}$ by considering both multiple muons and an angular range of $0^{\circ}\leq\theta\leq 90^{\circ}$. We estimate the fraction of muon bundles at the depth of SURF to be around 10\%. This is in contrast to the Homestake-measured fraction of 4.4\%~\cite{Cherry:1983dp}. If all calculations were corrected consistently for muon bundles, we may see better agreement.

Additionally, despite all measurements being taken at the 4850-level of the mine, another factor is the variation in elevation of nearly 0.57~km.w.e.\ in the overburden above SURF (see Fig.\ 5 in Ref.~\cite{MAJORANA:2016ifg}), as well as the lateral separation in locations between the Homestake and \mbox{MAJORANA} detectors, which, as discussed, can significantly affect \mute{} results.

\subsubsection{SNOLAB}
Although agreement can be achieved between the \mute{} prediction and the SNO data point from Ref.~\cite{SNO:2009oor} when all systematic uncertainties are considered, it should be noted that this data point is not well understood, and, like for other laboratories, agreement is dependent on the depth we consider for the lab. Despite some ambiguity about the correct depth to use for SNOLAB, in our calculations, we take $h=5.920$~km.w.e.\ based on the depth $d=2.092$~km quoted in Ref.~\cite{SNO:2009oor}.

In general, \mute{} in combination with \daemonflux{} seems to overpredict the total muon flux at the depth of SNOLAB, which might be consistent with what we observe for the vertical intensity in Ref.~\cite{Fedynitch:2021ima} using \ddm{} and also what we observe for CJPL-I in \cref{fig:total_flux_ratio} at a similar depth. Because deep depths are associated with high surface energies, this may suggest that \daemonflux{} does not calculate muon fluxes at high energies with high accuracy. Agreement improves when considering other hadronic interaction models, as shown in \cref{fig:total_flux_ratio_other_models} in \cref{app:comparisons_to_additional_models}.

Because a description of SNO rock in terms of Sternheimer parameters is not found in the literature for a full simulation of the energy loss, and because Ref.~\cite{SNO:2009oor} provides the only published measurement of the total underground muon flux for SNOLAB, further investigation is not possible at the moment. However, future measurements of the total muon flux at SNOLAB may help resolve this issue.

\subsection{Total flux underwater}

\begin{figure}
    \centering
    \includegraphics[width=\columnwidth]{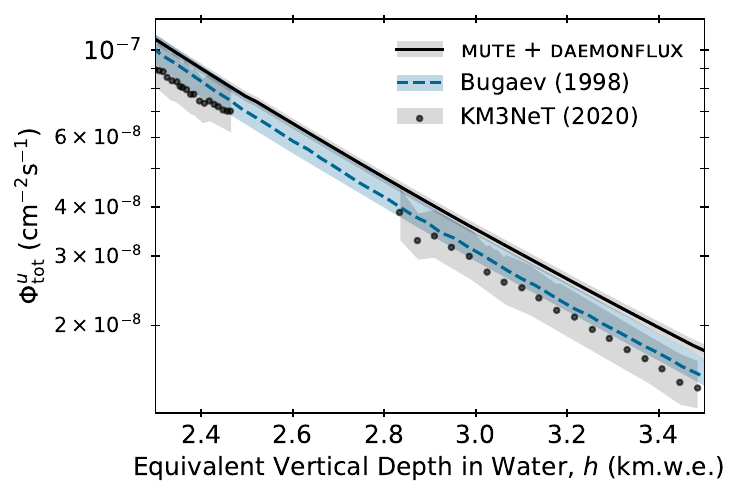}
    \caption{Total underwater muon flux vs slant depth. The \mute{} curve was calculated for a flat overburden using sea water (ANTARES water) as defined in \cref{tab:sternheimer} in \cref{app:rockcompositions}. A curve is also included for the theoretical calculation of Bugaev, et al.~\cite{Bugaev:1998bi}. KM3NeT data is taken from Ref.~\cite{KM3NeT:2019jfa}.}
    \label{fig:water_total_flux}
\end{figure}

Total muon flux measurements in water can be performed by water Cherenkov detectors, such as the KM3NeT detectors~\cite{KM3NeT:2019jfa}. The comparison between \mute{}, a typical reference calculation~\cite{Bugaev:1998bi}, and an early KM3NeT measurement using a single Detection Unit is shown in \cref{fig:water_total_flux}.

When performing computations for laboratories underwater as opposed to under a flat overburden, the only change in the \mute{} calculation is to the medium used for the surface-to-underground transfer tensor generation with \proposal{}, from rock to water. Thus, the accuracy of the \mute{} predictions in \cref{fig:total_flux_ratio} seen for rock should apply to water as well. We note that KM3NeT's data is a factor of 2 below our prediction for sea water. A more recent measurement of the event rate as a function of the zenith angle from a more complete version of the KM3NeT detector~\cite{KM3NeT:2024buf} indicates a 30--40\% discrepancy to a calculation made using a similar technique and the \sibyll{-2.3d} interaction model. Their result is approximately in agreement with our findings illustrated in \cref{fig:total_flux_ratio_other_models} (see \cref{app:comparisons_to_additional_models}) and also confirms that the previous results from Ref.~\cite{KM3NeT:2019jfa} shown in \cref{fig:water_total_flux} were significantly underestimated.

\section{\label{sec:angular_distributions}Angular Distributions}

We have used data from the LVD experiment located at LNGS to check the consistency of \mute{} predictions for the angular distribution of muons underground. The underground intensities from the LVD data, $I^u_{\mathrm{LVD}}$, were calculated according to

\begin{equation}
I^u_{\mathrm{LVD}}(\theta, \phi)=K\left(\frac{n(\theta, \phi)}{\varepsilon(\theta, \phi)}\right),
\end{equation}

\noindent where $K$ is a normalization constant representing the product of the detector lifetime and effective area to convert the $I^u_{\mathrm{LVD}}$ values to physical units, $n(\theta, \phi)$ is the number of raw counts, and $\varepsilon(\theta, \phi)$ is the acceptance of the LVD detector~\cite{LVD:1998lir}. $K$ was calculated by requiring the total integrated underground fluxes from \mute{}, $\Phi^u_{\mathrm{tot}}$, and from the LVD data to be equal:

\begin{equation}
\label{eq:K}
K=\frac{\Phi^u_{\mathrm{tot}}}{\iint_{\Omega}\left(\frac{n(\theta, \phi)}{\varepsilon(\theta, \phi)}\right)\,\mathrm{d}\Omega}.
\end{equation}

\begin{figure}
    \centering
    \includegraphics[width=\columnwidth]{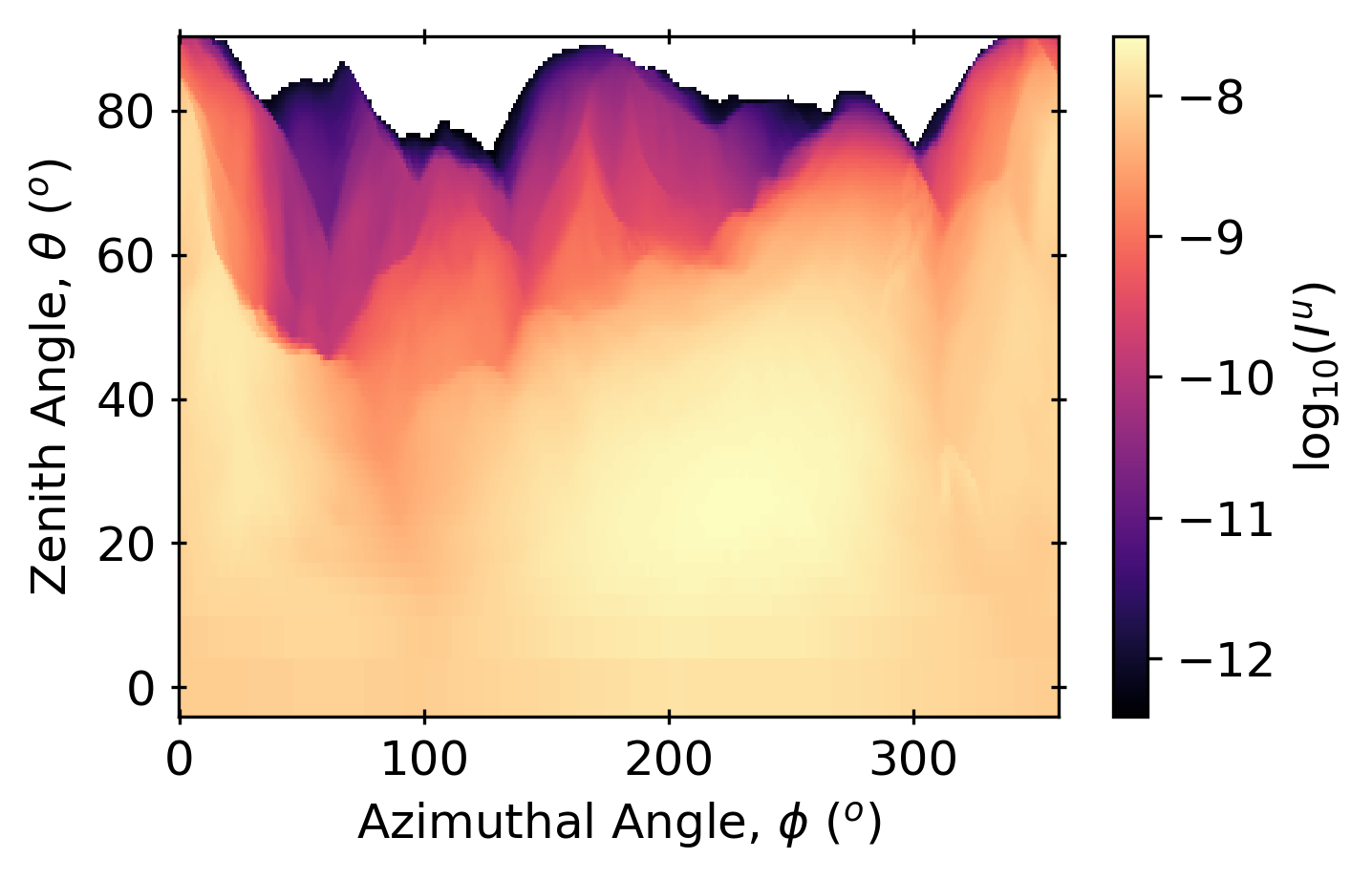}
    \includegraphics[width=\columnwidth]{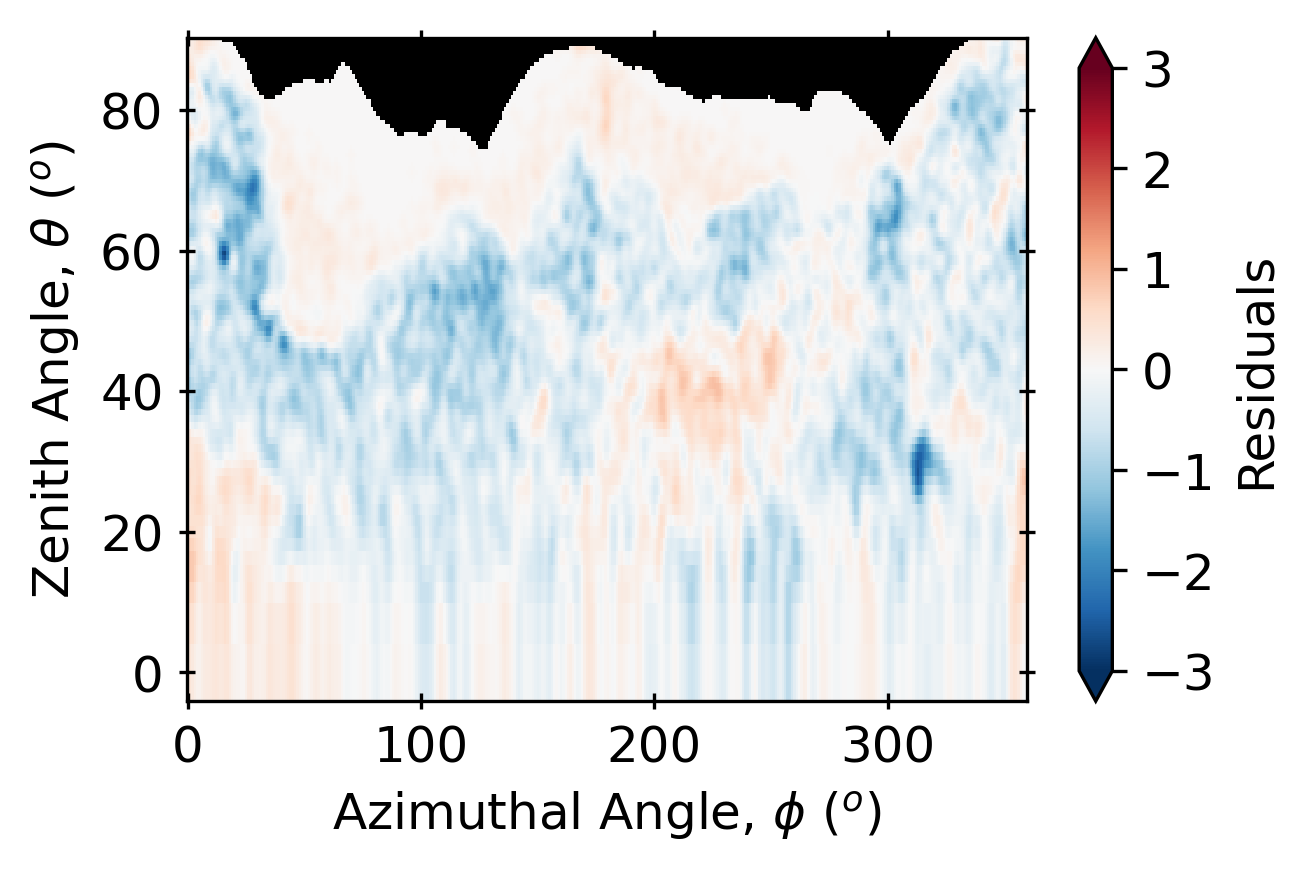}
    \caption{Underground double-differential muon intensities for the Gran Sasso mountain as calculated by \mute{} (top) and the residuals with respect to the LVD data (bottom) as a function of the zenith and azimuthal angles. In the figure of the residuals, black indicates masked bins for which the slant depth is greater than 14~km.w.e. Additionally, because the binning on the map of the Gran Sasso mountain is very fine, with large bin-to-bin fluctuations in the residuals, a Gaussian filter ($\sigma=1.5$) has been applied to help identify any characteristic trends.}
    \label{fig:lvdanalysis}
\end{figure}

\noindent The intensities were similarly computed using \mute{} with the method described in Sec. \ref{sec:computationalmethod}, and the results are shown in \cref{fig:lvdanalysis} (top). In order to compare the experimental and predicted intensities, the residuals for each $(\theta, \phi)$-bin of the data were calculated as

\begin{equation}
\mathrm{Residuals}=\frac{I^u_{\mathrm{LVD}}-I^u}{\delta I^u_{\mathrm{LVD}}},
\end{equation}

\noindent where $\delta I^u_{\mathrm{LVD}}$ are the statistical errors in each angular bin. The result is shown in \cref{fig:lvdanalysis} (bottom). Since, by construction from \cref{eq:K}, the normalizations of the \mute{} result and the LVD data match, only relative trends can be observed. The calculation is slightly below the data at near-vertical directions and above it for intermediate zenith angles, except for a small patch in the SW direction ($\sim225^{\circ}$ azimuth), where the situation is inverted. We calculate the $\chi^2/\mathrm{ndf}$ to be 1.34, where $\mathrm{ndf}$ is assumed to be the number of nonzero bins ($\mathrm{ndf}=31756$).

We first tested if we could obtain a better description of the data by allowing for a variable $E_{\mathrm{th}}$ in our computation of the intensities with \cref{eq:mountainintensity}. However, we found that the prediction and the data were independent of small variations in $E_{\mathrm{th}}$. We therefore investigated whether additional systematic uncertainty would be needed to accommodate this minor disagreement by exploring the expected surface azimuthal symmetry. We binned the observed intensities in slices of equal zenith angle and slant depth and compared the widths of the distributions between data and simulation. We expected to see the same flux within these bins with minor deviations coming from uncertainties in the data and calculations, which would result in narrow distributions. We found, however, that the distributions from the data were always wider than those from the simulation. We conclude, therefore, that the observed $\chi^2/\mathrm{ndf}$ is an effect of this additional systematic rather than an indication of a mismatch between model and data. Given this, the disagreement between \mute{} and LVD is not concerning for predicting the total flux.

The one-dimensional projections of the angular distributions in the zenith and azimuthal directions have also been calculated using

\begin{equation}
\label{eq:costheta_spectrum}
\Phi^u_{\phi}=\int I^u(X(\theta, \phi), \theta)\,\mathrm{d}\phi,
\end{equation}

\begin{equation}
\label{eq:phi_spectrum}
\Phi^u_{\theta}=\int I^u(X(\theta, \phi), \theta)\,\mathrm{d}\cos(\theta).
\end{equation}

\begin{figure}
    \centering
    \includegraphics[width=\columnwidth]{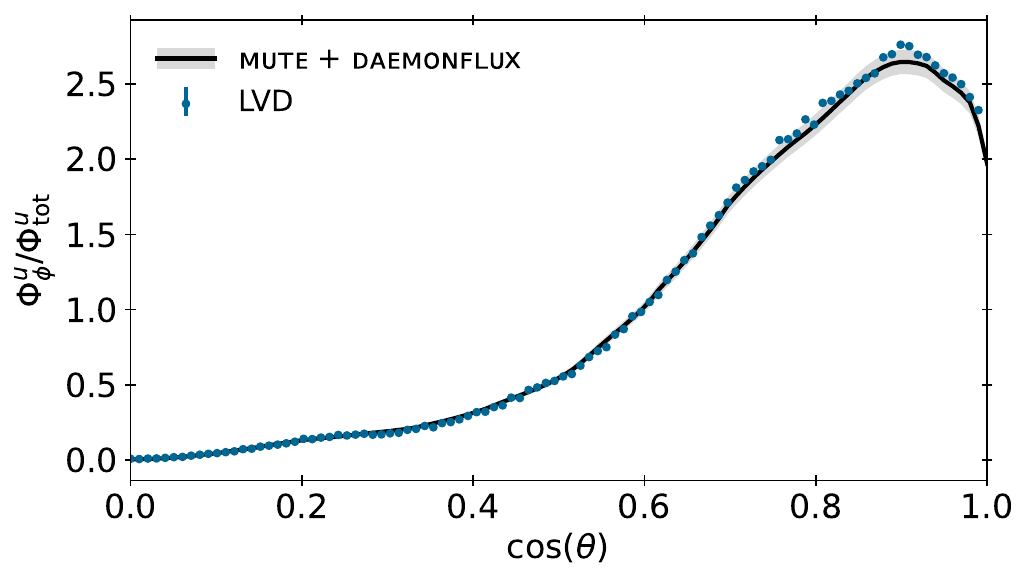}    \includegraphics[width=\columnwidth]{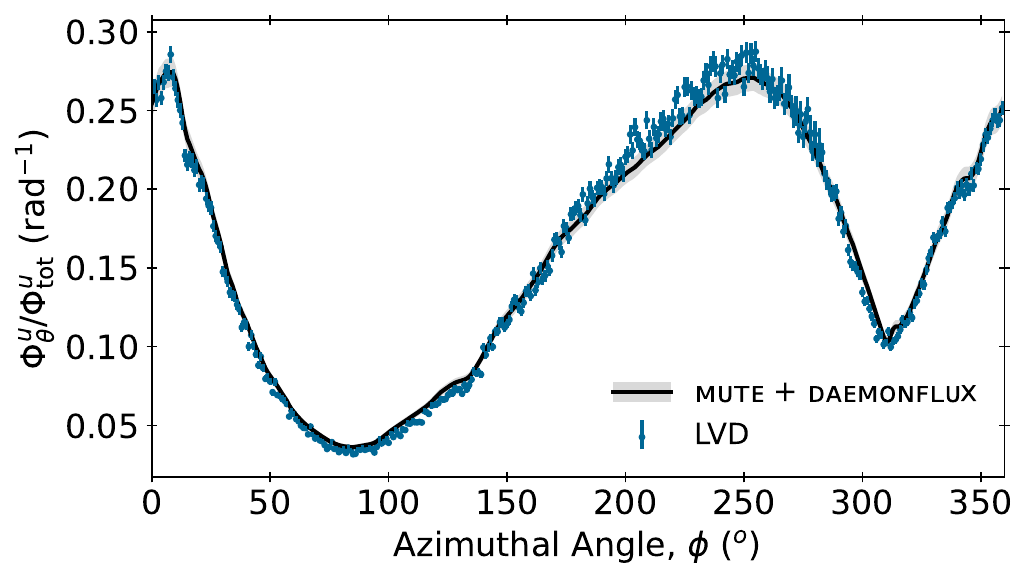}
    \caption{One-dimensional projections of the zenith (top) and azimuthal (bottom) angular distributions for the Gran Sasso mountain as calculated by \mute{}, compared to data from the LVD detector~\cite{LVD:1998lir}. The uncertainty bands on the \mute{} curves come from the error in the \daemonflux{} model. The uncertainties on the LVD points are derived from the acceptance and raw counts data. The total underground muon flux has normalized the spectra to mitigate uncertainties from the hadronic and primary models.}
    \label{fig:gransassoangulardistribution}
\end{figure}

\noindent Once the distribution in \cref{fig:lvdanalysis} (top) is projected onto the zenith and azimuth axes (\cref{fig:gransassoangulardistribution} top and bottom, respectively), we observe good compatibility of the result from \mute{} within the small errors of the data, in particular for the zenith projection at near-horizontal directions. The worse agreement in the patch of positive residuals in the $(\theta\sim 40^{\circ}, \phi\sim 225^{\circ})$ region in the bottom panel of \cref{fig:lvdanalysis} is also visible in the bottom panel of \cref{fig:gransassoangulardistribution}. This is an azimuthal range which contributes greatly to the flux, therefore affecting the $\chi^2/\mathrm{ndf}$, but the overall agreement we find between the \mute{} result and the LVD data is otherwise good.

In general, comparing the experimental and calculated angular distributions can reveal critical information about the accuracy of data analyses of underground experiments. The overall agreement obtained with the LVD experimental data suggests that \mute{} can be a powerful and helpful tool to inform on potential sources of errors and misreconstructions in data analyses for underground experiments, mainly due to the absence of statistical errors.

\section{\label{sec:underground_spectra_and_mean_energies}Underground Spectra and Mean Energies}

The muon energy spectrum for a specific underground lab, $\Phi^u_{\Omega}$, is defined by:

\begin{equation}
\label{eq:energy_spectrum}
\Phi_{\Omega}^u(E^u)=\int_{0}^{1}\int_{0}^{2\pi}\Phi^u(E^u, X(\theta, \phi), \theta)\mathrm{d}\cos(\theta)\mathrm{d}\phi.
\end{equation}

\begin{figure}
    \centering
    \includegraphics[width=\columnwidth]{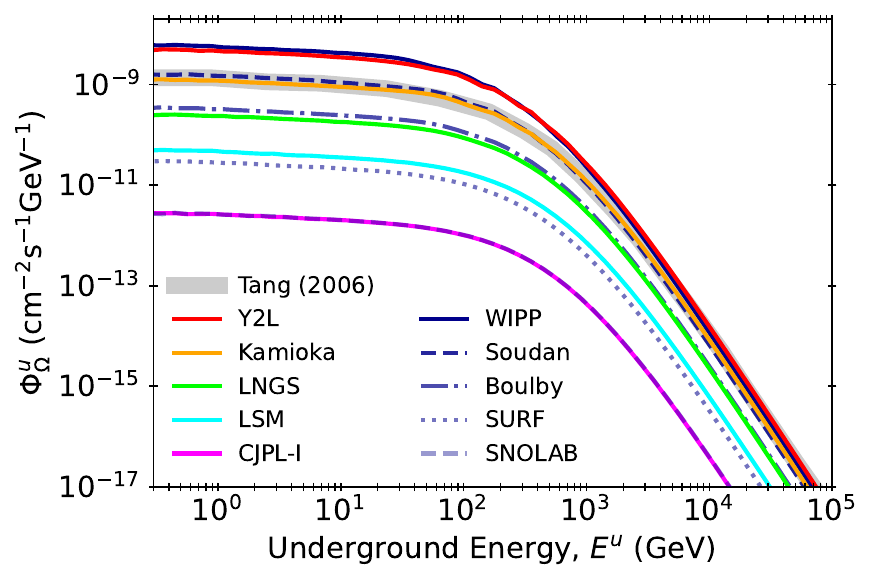}
    \caption{Underground muon energy spectra for labs under flat overburdens and mountains, calculated by \mute{} with \cref{eq:energy_spectrum}. The energy spectrum given in Ref.~\cite{Tang:2006uu} for Kamioka from a \music{}/\musun{} simulation is shown by the thick gray curve.}
    \label{fig:energy_spectra}
\end{figure}

\noindent In the case of labs under mountains, the zenith and azimuthal angles are those provided by the grid of the mountain map, whereas, for flat overburdens, \cref{eq:slant_depth} is used. Energy spectra calculated with \mute{} for labs under flat earth and under mountains are shown in \cref{fig:energy_spectra}. The curves' vertical ordering corresponds to the labs' depths, with some influence from the individual labs' rock compositions. In general, however, as expected, the energy spectrum decreases as depth increases. The spectra were verified by integrating the curves and comparing the results to the total underground muon flux for the given lab found with \cref{eq:total_flux}, which is calculated via a different computational method, and the results matched within 1\%.

Little data in physical units is available in the literature to compare our calculated energy spectra to, though there is a simulation result for Kamioka in Ref.~\cite{Tang:2006uu}, shown by the thick gray line in \cref{fig:energy_spectra}, which closely matches the \mute{} curve (the difference is less than 10\% for the majority of the displayed energy range).

The mean underground muon energy is also relevant to underground detectors, particularly in stopping vs through-going muons studies. Historically, it has been calculated by means of a parametrized energy spectrum,

\begin{equation}
\label{eq:meiandhimeenergyspectrum}
    \frac{\mathrm{d}N}{\mathrm{d}E^u}=Ae^{-bh(\gamma_{\mu}-1)}(E^u+\epsilon_{\mu}(1-e^{-bh}))^{-\gamma_{\mu}},
\end{equation}

\noindent where $A$ is a normalization constant with respect to the
differential muon intensity at a given depth $h$~\cite{Gaisser:2016uoy,ParticleDataGroup:2022pth, Mei:2005gm}. Values for the parameters $\epsilon_{\mu}$, $b$, and $\gamma_{\mu}$ for standard rock are listed in Ref.~\cite{Mei:2005gm} for the parametrizations from both Lipari, et al.~\cite{Lipari:1991ut} ($\epsilon_{\mu}=618~\mathrm{GeV}, b=0.383~\mathrm{km.w.e.}^{-1}, \gamma_{\mu}=3.7$), and Groom, et al.~\cite{Groom:2001kq,MACRO:1998zgn} ($\epsilon_{\mu}=693~\mathrm{GeV}, b=0.4~\mathrm{km.w.e.}^{-1}, \gamma_{\mu}=3.77$).

\begin{figure}
    \centering
    \includegraphics[width=\columnwidth]{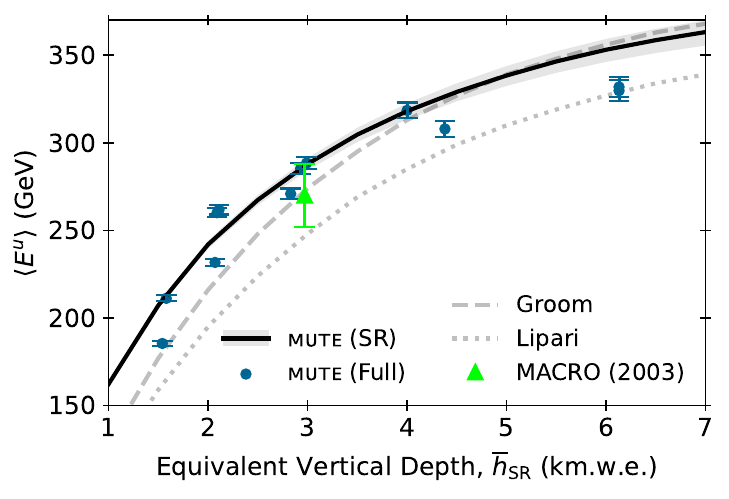}
    \caption{Mean underground energy vs vertical depth, showing values from \cref{tab:average_energies}. The \mute{} (Full) points are plotted at the \mute{}-inferred $\overline{h}_{\mathrm{SR}}$ depths given in \cref{tab:depthsandtotalfluxes} to be comparable to the \mute{} (SR) curve. The error bars on the \mute{} (Full) points and the error band on the \mute{} (SR) curve represent the uncertainty coming from the \daemonflux{} model. The MACRO point is a measurement for LNGS of $(270\pm 18)$~GeV from Ref.~\cite{MACRO:2002jmi}.}
    \label{fig:mean_energy}
\end{figure}

The mean energy is given by the first raw moment of the underground muon energy spectrum:

\begin{equation}
\label{eq:muteaverageenergy}
\langle E^u\rangle=\frac{\int_{0}^{\infty}E^u\Phi^u_{\Omega}(E^u)\,\mathrm{d}E^u}{\int_{0}^{\infty}\Phi^u_{\Omega}(E^u)\,\mathrm{d}E^u},
\end{equation}

\noindent where the integral in the denominator is the underground intensity, $I^u(X, \theta)$, from \cref{eq:mountainintensity}. We have done this calculation using the energy spectra from both \cref{eq:energy_spectrum,eq:meiandhimeenergyspectrum} for the underground sites listed in \cref{tab:undergroundlabs}, with the results given in \cref{tab:average_energies} and plotted in \cref{fig:mean_energy}. We have used the Lipari and Groom parameter sets with the $\overline{h}_{\mathrm{SR}}$ depths inferred from \mute{} given in \cref{tab:depthsandtotalfluxes}. Because these parameters are for standard rock, we also provide \mute{} results for standard rock at the $\overline{h}_{\mathrm{SR}}$ depths for comparison. Lastly, full \mute{} predictions are provided, using the rock compositions listed in \cref{tab:rock_compositions} and the labs' depths from \cref{tab:undergroundlabs} for flat overburden labs and the topographic maps for mountain labs.

\begin{table}[]
\caption{\label{tab:average_energies}Mean underground muon energies in GeV. The first three columns are comparable to each other, as they show results calculated with standard rock using the $\overline{h}_{\mathrm{SR}}$ values from \cref{tab:depthsandtotalfluxes} as depths. The last column provides our full calculation for the expected mean underground muon energy for each lab site, and uses the laboratory rocks as defined in \cref{tab:rock_compositions} as well as the depths $h$ from \cref{tab:undergroundlabs} for flat overburdens and the topographic maps for mountain labs. The uncertainties come from the error in the \daemonflux{} model.}
\begin{ruledtabular}
\begin{tabular}{@{}lccc|c@{}}
Laboratory & Lipari~\cite{Lipari:1991ut} & Groom~\cite{Groom:2001kq} & \mute{} (SR) & \mute{} (Full)\\
\colrule
WIPP & 162 & 180 & 210 & $185\pm 2$\\
Y2L & 165 & 183 & 213 & $211\pm 2$\\
Soudan & 199 & 221 & 245 & $232\pm 2$\\
Super-K & 200 & 222 & 246 & $260\pm 3$\\
KamLAND & 202 & 223 & 247 & $262\pm 3$\\
Boulby & 241 & 265 & 280 & $271\pm 3$\\
SUPL & 245 & 270 & 284 & $285\pm 3$\\
LNGS & 248 & 273 & 287 & $288\pm 4$\\
LSM & 285 & 312 & 316 & $318\pm 5$\\
SURF & 296 & 324 & 325 & $308\pm 5$\\
SNOLAB & 329 & 358 & 350 & $332\pm 7$\\
CJPL-I & 329 & 358 & 350 & $330\pm 7$\\
\end{tabular}
\end{ruledtabular}
\end{table}

Overall, within the widths of the distributions, the results are compatible with each other. Additionally, we find agreement between our calculation for LNGS using the laboratory rock and the MACRO measurement from Ref.~\cite{MACRO:2002jmi}.

\section{Conclusion}
We have presented a comprehensive compilation of the latest available references related to underground and underwater muon flux studies, and have compared to results from the muon intensity code \mute{} to explore the physics of cosmic-ray muons. The latest version of our open-source code, \mute{} v3, offers the ability to compute underground energy spectra and angular distributions as well as the ability to set custom rock compositions. It furthermore expands beyond calculations for labs under flat overburdens by now offering all calculations for sea level and labs under mountains as well. As it is both fast and flexible, \mute{} can be a valuable tool for laboratories to efficiently study their overburdens and data with little computational strain.

Using the data-driven \daemonflux{} model, we achieve a remarkable accuracy, particularly at depths below 5~km.w.e., confirming consistency between surface and underground flux measurements. Moreover, our findings indicate that hadronic interaction models like \sibyll{-2.3d} underpredict muon fluxes in the TeV surface energy range, something that has also been observed by underwater measurements from the KM3NeT detector.

Our findings highlight the importance of both the density and the chemical composition of the rock overburden above underground labs. Detailed knowledge of these is crucial in order to achieve an accuracy of significantly below 15\% in underground muon flux studies, and it is insufficient to use standard rock or parametric formulas. Therefore, we encourage experiments to study and publish details on the rock above their labs, namely average rock densities with uncertainties, $\left<Z/A\right>$ and $\left<Z^2/A\right>$ values (or minor or major component geochemical measurements), and Sternheimer parameters. Throughout our work, we also found that many laboratories have varying reported vertical depths in the literature. To achieve high-accuracy results, we find that it is useful to have access to published topographic maps of overburdens in terms of $X(\theta, \phi)$ in km.w.e., rather than only an average density, even for those labs under relatively flat earth. Additionally, it is helpful to publish intensity and flux data that has not been corrected to standard rock, allowing for external analyses to apply their own corrections in a consistent and transparent way. Lastly, we encourage laboratories to publish more detailed muon flux measurements, systematic uncertainties, and angular distributions whenever possible, even for decommissioned detectors.

This work has demonstrated that \mute{} is a suitable tool for interpreting underground muon measurements. It also made clear the importance of accurate modeling of the characteristics of each laboratory's overburden in order to achieve sufficiently high precision for indirect studies of muon and neutrino flux uncertainties in the TeV to PeV range at the surface.

\begin{acknowledgments}
We acknowledge the help of Marco Selvi and the LVD collaboration, who provided us with the topographic map of the Gran Sasso mountain and the LVD data for the muon angular distribution. We acknowledge, as well, the help of Michel Zampaolo and Luigi Mosca for the data of the Fréjus detector, and Holger Kluck and Wolfgang Rhode for a map of the Fréjus mountain. We also acknowledge Eunju Jeon and Hyun Su Lee, who provided us with a topographic map of the Yangyang Mountain and discussed with us muon bundles in the COSINE-100 detector, Itaru Shimizu and Shigetaka Moriyama for the maps of the Kamioka Mountain, and Shaomin Chen for the map of the Jinping Mountain. We want to thank Joseph Formaggio, Christopher Kyba, and Chris Waltham for their helpful discussions about the SNO detector and measurement. In addition, we acknowledge Jaret Heise and Kyle Jankord for providing the geochemical information for SURF rock formations and a three-dimensional SURF geology model, as well as the help of Joshua Albert for information on the overburden of WIPP. This research was enabled in part by support provided by the Digital Research Alliance of Canada and the Academia Sinica Grid-Computing Center (Grant No. AS-CFII-112-103). W.\ Woodley and M.-C.\ Piro acknowledge the support from the Canada First Research Excellence Fund through the Arthur B. McDonald Canadian Astroparticle Physics Research Institute.
\end{acknowledgments}

\appendix

\section{VARIABLE DEFINITIONS}
\label{app:variabledefinitions}
We provide here the explicit definitions of the variables used throughout the text to clarify the notation convention we use for fluxes and intensities.

The flux ($\Phi$, at the surface or underground) is defined as the number of particles ($\mathrm{d}N$) per unit area ($\mathrm{d}A$) per unit time ($\mathrm{d}t$) per solid angle ($\mathrm{d}\Omega$) per energy bin ($\mathrm{d}E$):

\begin{equation}
\label{eq:generalflux}
\Phi=\frac{\mathrm{d}N}{\mathrm{d}A\mathrm{d}t\mathrm{d}\Omega\mathrm{d}E}.
\end{equation}

\noindent The term ``intensity," represented by $I$, is used solely to refer to the flux integrated over only energy:

\begin{equation}
\label{eq:generalintensity}
I=\int\Phi\,\mathrm{d}E=\frac{\mathrm{d}N}{\mathrm{d}A\mathrm{d}t\mathrm{d}\Omega}.
\end{equation}

\noindent The energy and angular distributions are defined as the flux from \cref{eq:generalflux} and the intensity from \cref{eq:generalintensity}, respectively, integrated over the angles:

\begin{equation}
\Phi_{\Omega}=\int\Phi\,\mathrm{d}\Omega=\frac{\mathrm{d}N}{\mathrm{d}A\mathrm{d}t\mathrm{d}E},
\end{equation}

\begin{equation}
\Phi_{\phi}=\int I\,\mathrm{d}\phi=\frac{\mathrm{d}N}{\mathrm{d}A\mathrm{d}t\mathrm{d}\cos(\theta)},
\end{equation}

\begin{equation}
\Phi_{\theta}=\int I\,\mathrm{d}\cos(\theta)=\frac{\mathrm{d}N}{\mathrm{d}A\mathrm{d}t\mathrm{d}\phi}.
\end{equation}

\noindent Note that the subscript refers to the variable that has been integrated over, not the variable that the quantity is differential with respect to. Lastly, the total flux is defined as the number of muons per unit area per unit time:

\begin{equation}
\Phi_{\mathrm{tot}}=\int I\,\mathrm{d}\Omega=\frac{\mathrm{d}N}{\mathrm{d}A\mathrm{d}t}.
\end{equation}

\section{VALUES USED TO MODEL LABORATORY ROCK TYPES}
\label{app:rockcompositions}
As described in Secs. \ref{sec:computationalmethod} and \ref{sec:rock_composition}, rock compositions and Sternheimer parameters for different media are used in our \proposal{} simulations to calculate energy losses in \mute{} v3. \Cref{tab:geochemical,tab:sternheimer} report the minor component chemical compositions and Sternheimer parameters, respectively, which are used to obtain the main results in our work. The $\left<Z/A\right>$ and $\left<Z^2/A\right>$ values are defined as:

\begin{equation}
\left<\frac{Z}{A}\right>=\sum_if_i\frac{Z_i}{A_i},
\end{equation}

\begin{equation}
\left<\frac{Z^2}{A}\right>=\sum_if_i\frac{Z^2_i}{A_i},
\end{equation}

\noindent where $i$ is the $i$th element of the medium and $f$ is the mass fraction~\cite{menon1967progress}.

\begin{table*}[]
\caption{\label{tab:geochemical}Percent weights (\%) of minor component rock compositions for laboratories considered in this study, where available. Rock compositions were unavailable for WIPP, Boulby, and SUPL. In some cases, percent weights might not sum to 100\%, as some minor components that do not contribute significantly to the makeup of the rock have been excluded from reporting.}
\begin{ruledtabular}
\begin{tabular}{@{}lllllllllllr@{}}
Laboratory & H & C & O & Na & Mg & Al & Si & K & Ca & Fe & Reference\\
\colrule
$Z$ & 1 & 6 & 8 & 11 & 12 & 13 & 14 & 19 & 20 & 26 & \\
$A$ & 1 & 12 & 16 & 23 & 24 & 27 & 28 & 39 & 40 & 56 & \\
\colrule
Y2L & 0 & 0 & 48.00 & 2.60 & 2.40 & 8.40 & 26.00 & 1.90 & 4.20 & 5.20 & \cite{Yoon:2021tkv}\\
Soudan\footnote{Chemical compositions are available as reported here, but the $\left<Z/A\right>$ and $\left<Z^2/A\right>$ values from Ref.~\cite{ruddick1996underground} for Soudan and Ref.~\cite{FREJUS:1989lko} for LSM were used instead of those computed using this table.} & 0.30 & 0.08 & 44.89 & 1.85 & 392 & 7.94 & 23.65 & 0.33 & 6.43 & 9.34 & \cite{ruddick1996underground}\\
Kamioka\footnote{Multiple rock samples from Ikenoyama Mountain above the Kamioka lab are listed in Ref.~\cite{Mizukoshi:2018rnt}. Here we have used the average of the two JR-1 and JA-3 igneous samples, which are widely distributed around the Kamioka area, as given in Ref.~\cite{Yoon:2021tkv}.} & 0 & 0 & 39.00 & 0.01 & 0.60 & 6.00 & 17.00 & 0 & 28.00 & 7.60 & \cite{Yoon:2021tkv,Mizukoshi:2018rnt}\\
LNGS & 0.03 & 12.17 & 50.77 & 0 & 8.32 & 0.63 & 1.05 & 0.10 & 26.89 & 0 & \cite{MACRO:1995egd,Bussino:1994br}\\
LSM\footnotemark[1] & 0 & 0 & 24.00 & 0 & 5.60 & 1.00 & 1.30 & 0.10 & 30.00 & 0 & \cite{Yoon:2021tkv,Wulandari:2004bj}\\
SURF & 1.20 & 0 & 48.37 & 2.13 & 4.22 & 7.20 & 20.43 & 0.17 & 5.65 & 9.87 & \cite{Mei:2009py}\\
SNOLAB & 0.15 & 0.04 & 46.00 & 2.20 & 3.30 & 9.00 & 26.20 & 1.20 & 5.20 & 6.20 & \cite{Ewan:1987oqj}\\
CJPL & 0 & 9.59 & 46.42 & 0.01 & 11.50 & 0.15 & 0.19 & 0.07 & 31.96 & 0.10 & \cite{CDEX:2021cll}\\
\end{tabular}
\end{ruledtabular}
\end{table*}

\begin{table}[]
\caption{\label{tab:sternheimer}Sternheimer parameters for media used in this work. $I$ is the ionization energy of the medium, and the $\delta_0$ parameter from \cref{eq:sternheimerparameters} is 0 for all media. Values are taken from~\cite{koehne2013proposal}.}
\begin{ruledtabular}
\begin{tabular}{@{}lllllll@{}}
Medium & $I$ (eV) & $x_0$ & $x_1$ & $a$ & $b$ & $-c$\\
\colrule
Standard Rock & 136.4 & 0.0492 & 3.0549 & 0.08301 & 3.4120 & 3.7738\\
Fréjus Rock & 149.0 & 0.288 & 3.196 & 0.078 & 3.645 & 5.053\\
Salt & 175.3 & 0.2 & 3.0 & 0.1632 & 3 & 4.5041\\
Fresh Water & 79.7 & 0.2400 & 2.9004 & 0.09116 & 3.4773 & 3.5017\\
Sea Water & 75.0 & 0.2400 & 2.8004 & 0.09116 & 3.4773 & 3.5017\\
\end{tabular}
\end{ruledtabular}
\end{table}

The definitions of fresh water and sea water are very similar in terms of their Sternheimer parameters. However, their mass densities---given in Sec. \ref{sec:computationalmethod}---and their chemical compositions are different~\cite{koehne2013proposal}. These differences lead to total fluxes calculated with sea water being lower than those calculated with fresh water by about 5\%, constant across all depths.

\section{COMPARISONS TO ADDITIONAL HADRONIC INTERACTION MODELS}
\label{app:comparisons_to_additional_models}

\begin{figure*}[h]
    \centering
    \includegraphics[width=\columnwidth]{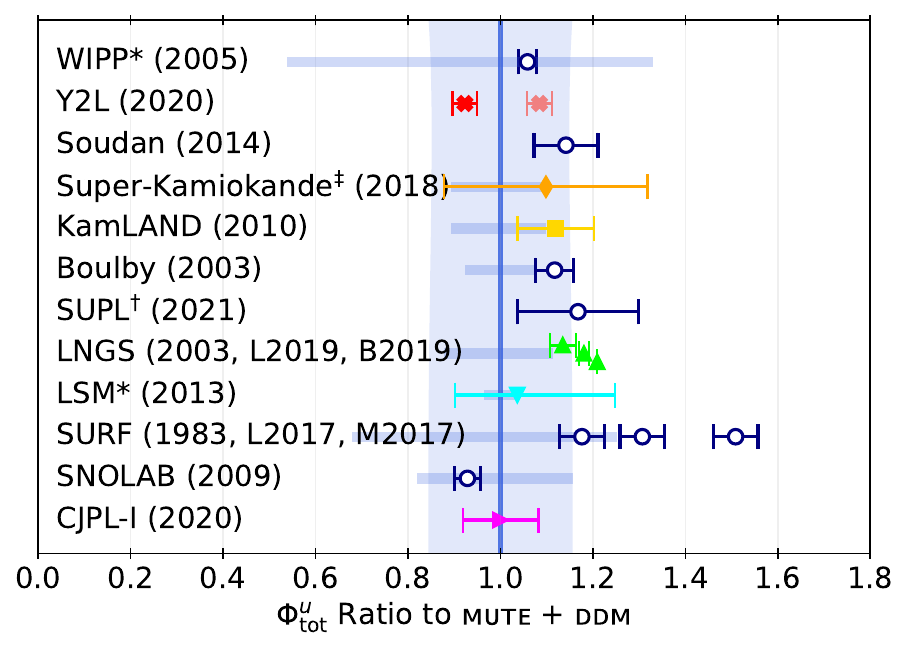}
    \includegraphics[width=\columnwidth]{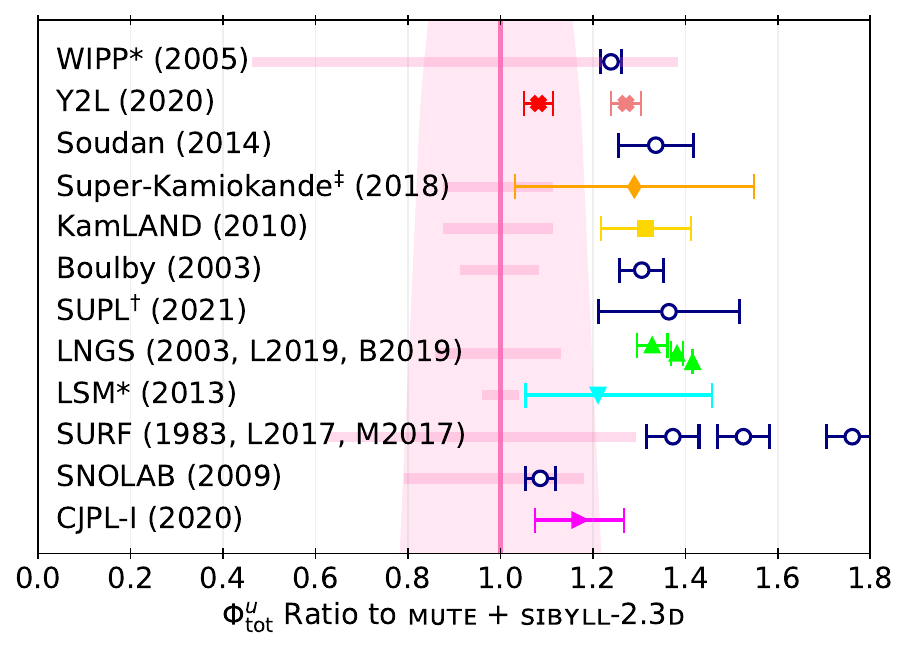}
    \includegraphics[width=\columnwidth]{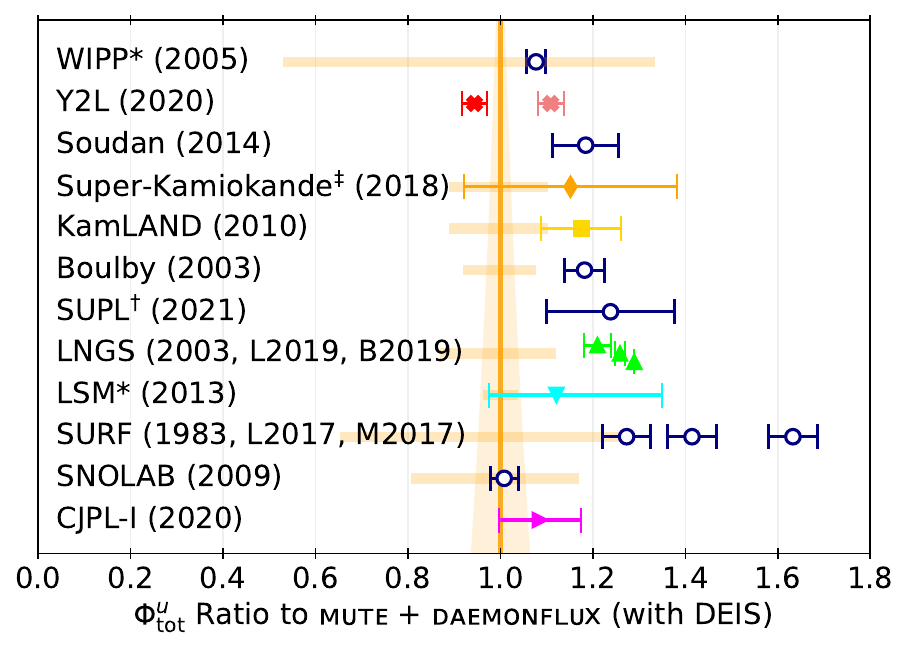}
    \includegraphics[width=\columnwidth]{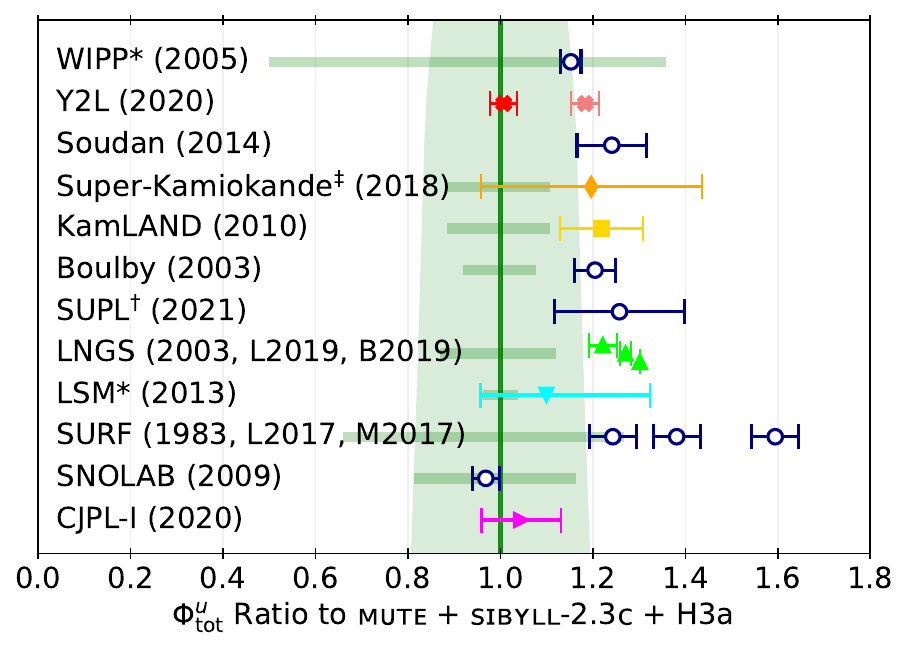}
    \caption{The ratio of experimental total underground muon flux measurements to \mute{} using the Data-Driven Model (\ddm{}; top left)~\cite{Fedynitch:2022vty}, \daemonflux{} calibrated including the DEIS dataset (bottom left)~\cite{Yanez:2023lsy}, \sibyll{-2.3d} (top right)~\cite{Riehn:2019jet}, and \sibyll{-2.3c}~\cite{Fedynitch:2018cbl} with the H3a primary flux model (bottom right)~\cite{Gaisser:2011klf}. The \ddm{} and \sibyll{-2.3d} panels use Global Spline Fit (\gsf{})~\cite{Dembinski:2017zsh} for the primary flux model. The error bands for the \sibyll{-2.3d} and \sibyll{-2.3c}+H3a panels come from the Bartol error scheme of Ref.~\cite{Barr:2006it}. For further explanation of the error bars, bands, and labels, refer to \cref{fig:total_flux_ratio}.}
    \label{fig:total_flux_ratio_other_models}
\end{figure*}

\Cref{fig:total_flux_ratio_other_models} shows the total flux ratio to \mute{} with different models of the surface flux. In Ref.~\cite{Fedynitch:2021ima}, we used \ddm{} as the baseline model for vertical equivalent flux calculations and obtained a result that was compatible with the data within errors. The central prediction had a tendency to underestimate fluxes at shallower depths (LVD and MACRO) and overshoot deeper data (LVD and SNO), both at the level of 5--10\%. The current result, shown in the upper left panel of \cref{fig:total_flux_ratio_other_models}, is consistent with this observation.

A dedicated \daemonflux{} fit, which included the near-horizontal muon flux measurement from DEIS (see Appendix A of Ref.~\cite{Yanez:2023lsy}), is shown in the lower left panel and confirms the initial suspicion in Ref.~\cite{Yanez:2023lsy} that the DEIS data may have unaccounted for systematic errors and therefore imposed an unnatural pull on the \daemonflux{} parameters.

The right panels of \cref{fig:total_flux_ratio_other_models} show the results using \sibyll{} hadronic interaction models, which consistently underestimate shallower total muon fluxes and are only consistent with the deepest measurements from SNO and CJPL. In addition to \gsf{}, a calculation is also shown for a different cosmic ray flux parameterization (H3a)~\cite{Gaisser:2011klf}, and while the tension with data is slightly reduced with this model, the model is still $\sim30\%$ lower than data. This result is consistent with the recent underwater measurement by KM3NeT~\cite{KM3NeT:2024buf}, where a simulation based on \gsf{} and \sibyll{-2.3d} is 30--40\% lower compared to the observed event rate. The impact of using a hadronic interaction model other than \sibyll{} is shown for the vertical equivalent intensities in Fig.~11 of Ref.~\cite{Fedynitch:2021ima} and can be expected to be similar in this work.

\bibliography{bibliography}

\begin{thebibliography}{98}%
\makeatletter
\providecommand \@ifxundefined [1]{%
 \@ifx{#1\undefined}
}%
\providecommand \@ifnum [1]{%
 \ifnum #1\expandafter \@firstoftwo
 \else \expandafter \@secondoftwo
 \fi
}%
\providecommand \@ifx [1]{%
 \ifx #1\expandafter \@firstoftwo
 \else \expandafter \@secondoftwo
 \fi
}%
\providecommand \natexlab [1]{#1}%
\providecommand \enquote  [1]{``#1''}%
\providecommand \bibnamefont  [1]{#1}%
\providecommand \bibfnamefont [1]{#1}%
\providecommand \citenamefont [1]{#1}%
\providecommand \href@noop [0]{\@secondoftwo}%
\providecommand \href [0]{\begingroup \@sanitize@url \@href}%
\providecommand \@href[1]{\@@startlink{#1}\@@href}%
\providecommand \@@href[1]{\endgroup#1\@@endlink}%
\providecommand \@sanitize@url [0]{\catcode `\\12\catcode `\$12\catcode `\&12\catcode `\#12\catcode `\^12\catcode `\_12\catcode `\%12\relax}%
\providecommand \@@startlink[1]{}%
\providecommand \@@endlink[0]{}%
\providecommand \url  [0]{\begingroup\@sanitize@url \@url }%
\providecommand \@url [1]{\endgroup\@href {#1}{\urlprefix }}%
\providecommand \urlprefix  [0]{URL }%
\providecommand \Eprint [0]{\href }%
\providecommand \doibase [0]{http://dx.doi.org/}%
\providecommand \selectlanguage [0]{\@gobble}%
\providecommand \bibinfo  [0]{\@secondoftwo}%
\providecommand \bibfield  [0]{\@secondoftwo}%
\providecommand \translation [1]{[#1]}%
\providecommand \BibitemOpen [0]{}%
\providecommand \bibitemStop [0]{}%
\providecommand \bibitemNoStop [0]{.\EOS\space}%
\providecommand \EOS [0]{\spacefactor3000\relax}%
\providecommand \BibitemShut  [1]{\csname bibitem#1\endcsname}%
\let\auto@bib@innerbib\@empty
\bibitem [{\citenamefont {Barrett}\ \emph {et~al.}(1952)\citenamefont {Barrett}, \citenamefont {Bollinger}, \citenamefont {Cocconi}, \citenamefont {Eisenberg},\ and\ \citenamefont {Greisen}}]{Barrett:1952woo}%
  \BibitemOpen
  \bibfield  {author} {\bibinfo {author} {\bibfnamefont {P.~H.}\ \bibnamefont {Barrett}}, \bibinfo {author} {\bibfnamefont {L.~M.}\ \bibnamefont {Bollinger}}, \bibinfo {author} {\bibfnamefont {G.}~\bibnamefont {Cocconi}}, \bibinfo {author} {\bibfnamefont {Y.}~\bibnamefont {Eisenberg}}, \ and\ \bibinfo {author} {\bibfnamefont {K.}~\bibnamefont {Greisen}},\ }\href {\doibase 10.1103/RevModPhys.24.133} {\bibfield  {journal} {\bibinfo  {journal} {Rev. Mod. Phys.}\ }\textbf {\bibinfo {volume} {24}},\ \bibinfo {pages} {133} (\bibinfo {year} {1952})}\BibitemShut {NoStop}%
\bibitem [{\citenamefont {Workman}\ \emph {et~al.}(2022)\citenamefont {Workman} \emph {et~al.}}]{ParticleDataGroup:2022pth}%
  \BibitemOpen
  \bibfield  {author} {\bibinfo {author} {\bibfnamefont {R.~L.}\ \bibnamefont {Workman}} \emph {et~al.} (\bibinfo {collaboration} {Particle Data Group}),\ }\href {\doibase 10.1093/ptep/ptac097} {\bibfield  {journal} {\bibinfo  {journal} {Prog. Theor. Exp. Phys.}\ }\textbf {\bibinfo {volume} {2022}},\ \bibinfo {pages} {083C01} (\bibinfo {year} {2022})}\BibitemShut {NoStop}%
\bibitem [{\citenamefont {Barr}\ \emph {et~al.}(2006)\citenamefont {Barr}, \citenamefont {Gaisser}, \citenamefont {Robbins},\ and\ \citenamefont {Stanev}}]{Barr:2006it}%
  \BibitemOpen
  \bibfield  {author} {\bibinfo {author} {\bibfnamefont {G.~D.}\ \bibnamefont {Barr}}, \bibinfo {author} {\bibfnamefont {T.~K.}\ \bibnamefont {Gaisser}}, \bibinfo {author} {\bibfnamefont {S.}~\bibnamefont {Robbins}}, \ and\ \bibinfo {author} {\bibfnamefont {T.}~\bibnamefont {Stanev}},\ }\href {\doibase 10.1103/PhysRevD.74.094009} {\bibfield  {journal} {\bibinfo  {journal} {Phys. Rev. D}\ }\textbf {\bibinfo {volume} {74}},\ \bibinfo {pages} {094009} (\bibinfo {year} {2006})}\BibitemShut {NoStop}%
\bibitem [{\citenamefont {Honda}\ \emph {et~al.}(2007)\citenamefont {Honda}, \citenamefont {Kajita}, \citenamefont {Kasahara}, \citenamefont {Midorikawa},\ and\ \citenamefont {Sanuki}}]{Honda:2006qj}%
  \BibitemOpen
  \bibfield  {author} {\bibinfo {author} {\bibfnamefont {M.}~\bibnamefont {Honda}}, \bibinfo {author} {\bibfnamefont {T.}~\bibnamefont {Kajita}}, \bibinfo {author} {\bibfnamefont {K.}~\bibnamefont {Kasahara}}, \bibinfo {author} {\bibfnamefont {S.}~\bibnamefont {Midorikawa}}, \ and\ \bibinfo {author} {\bibfnamefont {T.}~\bibnamefont {Sanuki}},\ }\href {\doibase 10.1103/PhysRevD.75.043006} {\bibfield  {journal} {\bibinfo  {journal} {Phys. Rev. D}\ }\textbf {\bibinfo {volume} {75}},\ \bibinfo {pages} {043006} (\bibinfo {year} {2007})}\BibitemShut {NoStop}%
\bibitem [{\citenamefont {Ya\~nez}\ and\ \citenamefont {Fedynitch}(2023)}]{Yanez:2023lsy}%
  \BibitemOpen
  \bibfield  {author} {\bibinfo {author} {\bibfnamefont {J.~P.}\ \bibnamefont {Ya\~nez}}\ and\ \bibinfo {author} {\bibfnamefont {A.}~\bibnamefont {Fedynitch}},\ }\href {\doibase 10.1103/PhysRevD.107.123037} {\bibfield  {journal} {\bibinfo  {journal} {Phys. Rev. D}\ }\textbf {\bibinfo {volume} {107}},\ \bibinfo {pages} {123037} (\bibinfo {year} {2023})}\BibitemShut {NoStop}%
\bibitem [{\citenamefont {Formaggio}\ and\ \citenamefont {Martoff}(2004)}]{Formaggio:2004mm}%
  \BibitemOpen
  \bibfield  {author} {\bibinfo {author} {\bibfnamefont {J.~A.}\ \bibnamefont {Formaggio}}\ and\ \bibinfo {author} {\bibfnamefont {C.}~\bibnamefont {Martoff}},\ }\href {\doibase 10.1146/annurev.nucl.54.070103.181248} {\bibfield  {journal} {\bibinfo  {journal} {Ann. Rev. of Nucl. and Part. Sci.}\ }\textbf {\bibinfo {volume} {54}},\ \bibinfo {pages} {361} (\bibinfo {year} {2004})}\BibitemShut {NoStop}%
\bibitem [{\citenamefont {Bugaev}\ \emph {et~al.}(1998)\citenamefont {Bugaev}, \citenamefont {Misaki}, \citenamefont {Naumov}, \citenamefont {Sinegovskaya}, \citenamefont {Sinegovsky},\ and\ \citenamefont {Takahashi}}]{Bugaev:1998bi}%
  \BibitemOpen
  \bibfield  {author} {\bibinfo {author} {\bibfnamefont {E.~V.}\ \bibnamefont {Bugaev}}, \bibinfo {author} {\bibfnamefont {A.}~\bibnamefont {Misaki}}, \bibinfo {author} {\bibfnamefont {V.~A.}\ \bibnamefont {Naumov}}, \bibinfo {author} {\bibfnamefont {T.~S.}\ \bibnamefont {Sinegovskaya}}, \bibinfo {author} {\bibfnamefont {S.~I.}\ \bibnamefont {Sinegovsky}}, \ and\ \bibinfo {author} {\bibfnamefont {N.}~\bibnamefont {Takahashi}},\ }\href {\doibase 10.1103/PhysRevD.58.054001} {\bibfield  {journal} {\bibinfo  {journal} {Phys. Rev. D}\ }\textbf {\bibinfo {volume} {58}},\ \bibinfo {pages} {054001} (\bibinfo {year} {1998})}\BibitemShut {NoStop}%
\bibitem [{\citenamefont {Lipari}\ and\ \citenamefont {Stanev}(1991)}]{Lipari:1991ut}%
  \BibitemOpen
  \bibfield  {author} {\bibinfo {author} {\bibfnamefont {P.}~\bibnamefont {Lipari}}\ and\ \bibinfo {author} {\bibfnamefont {T.}~\bibnamefont {Stanev}},\ }\href {\doibase 10.1103/PhysRevD.44.3543} {\bibfield  {journal} {\bibinfo  {journal} {Phys. Rev. D}\ }\textbf {\bibinfo {volume} {44}},\ \bibinfo {pages} {3543} (\bibinfo {year} {1991})}\BibitemShut {NoStop}%
\bibitem [{\citenamefont {Crouch}(1987)}]{crouch1987improved}%
  \BibitemOpen
  \bibfield  {author} {\bibinfo {author} {\bibfnamefont {M.}~\bibnamefont {Crouch}},\ }in\ \href@noop {} {\emph {\bibinfo {booktitle} {Proceedings of the 20th International Cosmic Ray Conference Moscow}}},\ Vol.~\bibinfo {volume} {6}\ (\bibinfo  {publisher} {Nauka, Moscow},\ \bibinfo {year} {1987})\ p.\ \bibinfo {pages} {165}\BibitemShut {NoStop}%
\bibitem [{\citenamefont {Mei}\ and\ \citenamefont {Hime}(2006)}]{Mei:2005gm}%
  \BibitemOpen
  \bibfield  {author} {\bibinfo {author} {\bibfnamefont {D.}~\bibnamefont {Mei}}\ and\ \bibinfo {author} {\bibfnamefont {A.}~\bibnamefont {Hime}},\ }\href {\doibase 10.1103/PhysRevD.73.053004} {\bibfield  {journal} {\bibinfo  {journal} {Phys. Rev. D}\ }\textbf {\bibinfo {volume} {73}},\ \bibinfo {pages} {053004} (\bibinfo {year} {2006})}\BibitemShut {NoStop}%
\bibitem [{\citenamefont {Sokalski}\ \emph {et~al.}(2001)\citenamefont {Sokalski}, \citenamefont {Bugaev},\ and\ \citenamefont {Klimushin}}]{Sokalski:2000nb}%
  \BibitemOpen
  \bibfield  {author} {\bibinfo {author} {\bibfnamefont {I.~A.}\ \bibnamefont {Sokalski}}, \bibinfo {author} {\bibfnamefont {E.~V.}\ \bibnamefont {Bugaev}}, \ and\ \bibinfo {author} {\bibfnamefont {S.~I.}\ \bibnamefont {Klimushin}},\ }\href {\doibase 10.1103/PhysRevD.64.074015} {\bibfield  {journal} {\bibinfo  {journal} {Phys. Rev. D}\ }\textbf {\bibinfo {volume} {64}},\ \bibinfo {pages} {074015} (\bibinfo {year} {2001})}\BibitemShut {NoStop}%
\bibitem [{\citenamefont {Kudryavtsev}(2009)}]{Kudryavtsev:2008qh}%
  \BibitemOpen
  \bibfield  {author} {\bibinfo {author} {\bibfnamefont {V.~A.}\ \bibnamefont {Kudryavtsev}},\ }\href {\doibase 10.1016/j.cpc.2008.10.013} {\bibfield  {journal} {\bibinfo  {journal} {Comput. Phys. Commun.}\ }\textbf {\bibinfo {volume} {180}},\ \bibinfo {pages} {339} (\bibinfo {year} {2009})}\BibitemShut {NoStop}%
\bibitem [{\citenamefont {Agostinelli}\ \emph {et~al.}(2003)\citenamefont {Agostinelli} \emph {et~al.}}]{GEANT4:2002zbu}%
  \BibitemOpen
  \bibfield  {author} {\bibinfo {author} {\bibfnamefont {S.}~\bibnamefont {Agostinelli}} \emph {et~al.} (\bibinfo {collaboration} {GEANT4 Collaboration}),\ }\href {\doibase 10.1016/S0168-9002(03)01368-8} {\bibfield  {journal} {\bibinfo  {journal} {Nucl. Instrum. Methods Phys. Res., Sect. A}\ }\textbf {\bibinfo {volume} {506}},\ \bibinfo {pages} {250} (\bibinfo {year} {2003})}\BibitemShut {NoStop}%
\bibitem [{\citenamefont {Battistoni}\ \emph {et~al.}(2015)\citenamefont {Battistoni} \emph {et~al.}}]{Battistoni:2015epi}%
  \BibitemOpen
  \bibfield  {author} {\bibinfo {author} {\bibfnamefont {G.}~\bibnamefont {Battistoni}} \emph {et~al.},\ }\href {\doibase 10.1016/j.anucene.2014.11.007} {\bibfield  {journal} {\bibinfo  {journal} {Ann. Nucl. Energy}\ }\textbf {\bibinfo {volume} {82}},\ \bibinfo {pages} {10} (\bibinfo {year} {2015})}\BibitemShut {NoStop}%
\bibitem [{\citenamefont {Fedynitch}\ \emph {et~al.}(2022)\citenamefont {Fedynitch}, \citenamefont {Woodley},\ and\ \citenamefont {Piro}}]{Fedynitch:2021ima}%
  \BibitemOpen
  \bibfield  {author} {\bibinfo {author} {\bibfnamefont {A.}~\bibnamefont {Fedynitch}}, \bibinfo {author} {\bibfnamefont {W.}~\bibnamefont {Woodley}}, \ and\ \bibinfo {author} {\bibfnamefont {M.-C.}\ \bibnamefont {Piro}},\ }\href {\doibase 10.3847/1538-4357/ac5027} {\bibfield  {journal} {\bibinfo  {journal} {Astrophys. J.}\ }\textbf {\bibinfo {volume} {928}},\ \bibinfo {pages} {27} (\bibinfo {year} {2022})}\BibitemShut {NoStop}%
\bibitem [{\citenamefont {Fedynitch}(2015)}]{Fedynitch:2015kcn}%
  \BibitemOpen
  \bibfield  {author} {\bibinfo {author} {\bibfnamefont {A.}~\bibnamefont {Fedynitch}},\ }\emph {\bibinfo {title} {{Cascade equations and hadronic interactions at very high energies}}},\ \href {\doibase 10.5445/IR/1000055433} {Ph.D. thesis},\ \bibinfo  {school} {KIT, Karlsruhe, Dept. Phys.} (\bibinfo {year} {2015})\BibitemShut {NoStop}%
\bibitem [{\citenamefont {Fedynitch}\ \emph {et~al.}(2019)\citenamefont {Fedynitch}, \citenamefont {Riehn}, \citenamefont {Engel}, \citenamefont {Gaisser},\ and\ \citenamefont {Stanev}}]{Fedynitch:2018cbl}%
  \BibitemOpen
  \bibfield  {author} {\bibinfo {author} {\bibfnamefont {A.}~\bibnamefont {Fedynitch}}, \bibinfo {author} {\bibfnamefont {F.}~\bibnamefont {Riehn}}, \bibinfo {author} {\bibfnamefont {R.}~\bibnamefont {Engel}}, \bibinfo {author} {\bibfnamefont {T.~K.}\ \bibnamefont {Gaisser}}, \ and\ \bibinfo {author} {\bibfnamefont {T.}~\bibnamefont {Stanev}},\ }\href {\doibase 10.1103/PhysRevD.100.103018} {\bibfield  {journal} {\bibinfo  {journal} {Phys. Rev. D}\ }\textbf {\bibinfo {volume} {100}},\ \bibinfo {pages} {103018} (\bibinfo {year} {2019})}\BibitemShut {NoStop}%
\bibitem [{\citenamefont {Koehne}\ \emph {et~al.}(2013)\citenamefont {Koehne}, \citenamefont {Frantzen}, \citenamefont {Schmitz}, \citenamefont {Fuchs}, \citenamefont {Rhode}, \citenamefont {Chirkin},\ and\ \citenamefont {Tjus}}]{koehne2013proposal}%
  \BibitemOpen
  \bibfield  {author} {\bibinfo {author} {\bibfnamefont {J.-H.}\ \bibnamefont {Koehne}}, \bibinfo {author} {\bibfnamefont {K.}~\bibnamefont {Frantzen}}, \bibinfo {author} {\bibfnamefont {M.}~\bibnamefont {Schmitz}}, \bibinfo {author} {\bibfnamefont {T.}~\bibnamefont {Fuchs}}, \bibinfo {author} {\bibfnamefont {W.}~\bibnamefont {Rhode}}, \bibinfo {author} {\bibfnamefont {D.}~\bibnamefont {Chirkin}}, \ and\ \bibinfo {author} {\bibfnamefont {J.~B.}\ \bibnamefont {Tjus}},\ }\href {\doibase 10.1016/j.cpc.2013.04.001} {\bibfield  {journal} {\bibinfo  {journal} {Comput. Phys. Commun.}\ }\textbf {\bibinfo {volume} {184}},\ \bibinfo {pages} {2070} (\bibinfo {year} {2013})}\BibitemShut {NoStop}%
\bibitem [{\citenamefont {Dembinski}\ \emph {et~al.}(2018)\citenamefont {Dembinski}, \citenamefont {Engel}, \citenamefont {Fedynitch}, \citenamefont {Gaisser}, \citenamefont {Riehn},\ and\ \citenamefont {Stanev}}]{Dembinski:2017zsh}%
  \BibitemOpen
  \bibfield  {author} {\bibinfo {author} {\bibfnamefont {H.~P.}\ \bibnamefont {Dembinski}}, \bibinfo {author} {\bibfnamefont {R.}~\bibnamefont {Engel}}, \bibinfo {author} {\bibfnamefont {A.}~\bibnamefont {Fedynitch}}, \bibinfo {author} {\bibfnamefont {T.}~\bibnamefont {Gaisser}}, \bibinfo {author} {\bibfnamefont {F.}~\bibnamefont {Riehn}}, \ and\ \bibinfo {author} {\bibfnamefont {T.}~\bibnamefont {Stanev}},\ }\href {\doibase 10.22323/1.301.0533} {\bibfield  {journal} {\bibinfo  {journal} {Proc. Sci.}\ }\textbf {\bibinfo {volume} {ICRC2017}},\ \bibinfo {pages} {533} (\bibinfo {year} {2018})}\BibitemShut {NoStop}%
\bibitem [{\citenamefont {Fedynitch}\ and\ \citenamefont {Huber}(2022)}]{Fedynitch:2022vty}%
  \BibitemOpen
  \bibfield  {author} {\bibinfo {author} {\bibfnamefont {A.}~\bibnamefont {Fedynitch}}\ and\ \bibinfo {author} {\bibfnamefont {M.}~\bibnamefont {Huber}},\ }\href {\doibase 10.1103/PhysRevD.106.083018} {\bibfield  {journal} {\bibinfo  {journal} {Phys. Rev. D}\ }\textbf {\bibinfo {volume} {106}},\ \bibinfo {pages} {083018} (\bibinfo {year} {2022})}\BibitemShut {NoStop}%
\bibitem [{\citenamefont {Berger}\ \emph {et~al.}(1989)\citenamefont {Berger} \emph {et~al.}}]{FREJUS:1989lko}%
  \BibitemOpen
  \bibfield  {author} {\bibinfo {author} {\bibfnamefont {C.}~\bibnamefont {Berger}} \emph {et~al.} (\bibinfo {collaboration} {FREJUS Collaboration}),\ }\href {\doibase 10.1103/PhysRevD.40.2163} {\bibfield  {journal} {\bibinfo  {journal} {Phys. Rev. D}\ }\textbf {\bibinfo {volume} {40}},\ \bibinfo {pages} {2163} (\bibinfo {year} {1989})}\BibitemShut {NoStop}%
\bibitem [{\citenamefont {Ahlen}\ \emph {et~al.}(1990)\citenamefont {Ahlen} \emph {et~al.}}]{MACRO:1990ykz}%
  \BibitemOpen
  \bibfield  {author} {\bibinfo {author} {\bibfnamefont {S.~P.}\ \bibnamefont {Ahlen}} \emph {et~al.} (\bibinfo {collaboration} {MACRO Collaboration}),\ }\href {\doibase 10.1016/0370-2693(90)90541-D} {\bibfield  {journal} {\bibinfo  {journal} {Phys. Lett. B}\ }\textbf {\bibinfo {volume} {249}},\ \bibinfo {pages} {149} (\bibinfo {year} {1990})}\BibitemShut {NoStop}%
\bibitem [{\citenamefont {Ambrosio}\ \emph {et~al.}(1995)\citenamefont {Ambrosio} \emph {et~al.}}]{MACRO:1995egd}%
  \BibitemOpen
  \bibfield  {author} {\bibinfo {author} {\bibfnamefont {M.}~\bibnamefont {Ambrosio}} \emph {et~al.} (\bibinfo {collaboration} {MACRO Collaboration}),\ }\href {\doibase 10.1103/PhysRevD.52.3793} {\bibfield  {journal} {\bibinfo  {journal} {Phys. Rev. D}\ }\textbf {\bibinfo {volume} {52}},\ \bibinfo {pages} {3793} (\bibinfo {year} {1995})}\BibitemShut {NoStop}%
\bibitem [{\citenamefont {Aglietta}\ \emph {et~al.}(1998)\citenamefont {Aglietta} \emph {et~al.}}]{LVD:1998lir}%
  \BibitemOpen
  \bibfield  {author} {\bibinfo {author} {\bibfnamefont {M.}~\bibnamefont {Aglietta}} \emph {et~al.} (\bibinfo {collaboration} {LVD Collaboration}),\ }\href {\doibase 10.1103/PhysRevD.58.092005} {\bibfield  {journal} {\bibinfo  {journal} {Phys. Rev. D}\ }\textbf {\bibinfo {volume} {58}},\ \bibinfo {pages} {092005} (\bibinfo {year} {1998})}\BibitemShut {NoStop}%
\bibitem [{\citenamefont {Aharmim}\ \emph {et~al.}(2009)\citenamefont {Aharmim} \emph {et~al.}}]{SNO:2009oor}%
  \BibitemOpen
  \bibfield  {author} {\bibinfo {author} {\bibfnamefont {B.}~\bibnamefont {Aharmim}} \emph {et~al.} (\bibinfo {collaboration} {SNO Collaboration}),\ }\href {\doibase 10.1103/PhysRevD.80.012001} {\bibfield  {journal} {\bibinfo  {journal} {Phys. Rev. D}\ }\textbf {\bibinfo {volume} {80}},\ \bibinfo {pages} {012001} (\bibinfo {year} {2009})}\BibitemShut {NoStop}%
\bibitem [{\citenamefont {Woodley}\ and\ \citenamefont {Fedynitch}(2022)}]{william_woodley_2022_6841971}%
  \BibitemOpen
  \bibfield  {author} {\bibinfo {author} {\bibfnamefont {W.}~\bibnamefont {Woodley}}\ and\ \bibinfo {author} {\bibfnamefont {A.}~\bibnamefont {Fedynitch}},\ }\href {\doibase 10.5281/zenodo.6841971} {\enquote {\bibinfo {title} {wjwoodley/mute: Mute 2.0.0},}\ } (\bibinfo {year} {2022})\BibitemShut {NoStop}%
\bibitem [{\citenamefont {Abe}\ \emph {et~al.}(2010)\citenamefont {Abe} \emph {et~al.}}]{KamLAND:2009zwo}%
  \BibitemOpen
  \bibfield  {author} {\bibinfo {author} {\bibfnamefont {S.}~\bibnamefont {Abe}} \emph {et~al.} (\bibinfo {collaboration} {KamLAND Collaboration}),\ }\href {\doibase 10.1103/PhysRevC.81.025807} {\bibfield  {journal} {\bibinfo  {journal} {Phys. Rev. C}\ }\textbf {\bibinfo {volume} {81}},\ \bibinfo {pages} {025807} (\bibinfo {year} {2010})}\BibitemShut {NoStop}%
\bibitem [{\citenamefont {Fukuda}\ \emph {et~al.}(2003)\citenamefont {Fukuda} \emph {et~al.}}]{Super-Kamiokande:2002weg}%
  \BibitemOpen
  \bibfield  {author} {\bibinfo {author} {\bibfnamefont {Y.}~\bibnamefont {Fukuda}} \emph {et~al.} (\bibinfo {collaboration} {Super-Kamiokande Collaboration}),\ }\href {\doibase 10.1016/S0168-9002(03)00425-X} {\bibfield  {journal} {\bibinfo  {journal} {Nucl. Instrum. Methods Phys. Res., Sect. A}\ }\textbf {\bibinfo {volume} {501}},\ \bibinfo {pages} {418} (\bibinfo {year} {2003})}\BibitemShut {NoStop}%
\bibitem [{\citenamefont {Guo}\ \emph {et~al.}(2021)\citenamefont {Guo} \emph {et~al.}}]{JNE:2020bwn}%
  \BibitemOpen
  \bibfield  {author} {\bibinfo {author} {\bibfnamefont {Z.}~\bibnamefont {Guo}} \emph {et~al.} (\bibinfo {collaboration} {JNE Collaboration}),\ }\href {\doibase 10.1088/1674-1137/abccae} {\bibfield  {journal} {\bibinfo  {journal} {Chin. Phys. C}\ }\textbf {\bibinfo {volume} {45}},\ \bibinfo {pages} {025001} (\bibinfo {year} {2021})}\BibitemShut {NoStop}%
\bibitem [{\citenamefont {Esch}\ \emph {et~al.}(2005)\citenamefont {Esch}, \citenamefont {Bowles}, \citenamefont {Hime}, \citenamefont {Pichlmaier}, \citenamefont {Reifarth},\ and\ \citenamefont {Wollnik}}]{Esch:2004zj}%
  \BibitemOpen
  \bibfield  {author} {\bibinfo {author} {\bibfnamefont {E.-I.}\ \bibnamefont {Esch}}, \bibinfo {author} {\bibfnamefont {T.~J.}\ \bibnamefont {Bowles}}, \bibinfo {author} {\bibfnamefont {A.}~\bibnamefont {Hime}}, \bibinfo {author} {\bibfnamefont {A.}~\bibnamefont {Pichlmaier}}, \bibinfo {author} {\bibfnamefont {R.}~\bibnamefont {Reifarth}}, \ and\ \bibinfo {author} {\bibfnamefont {H.}~\bibnamefont {Wollnik}},\ }\href {\doibase 10.1016/j.nima.2004.09.005} {\bibfield  {journal} {\bibinfo  {journal} {Nucl. Instrum. Meth. A}\ }\textbf {\bibinfo {volume} {538}},\ \bibinfo {pages} {516} (\bibinfo {year} {2005})}\BibitemShut {NoStop}%
\bibitem [{\citenamefont {Prihtiadi}(2018)}]{Prihtiadi:2017sxc}%
  \BibitemOpen
  \bibfield  {author} {\bibinfo {author} {\bibfnamefont {H.}~\bibnamefont {Prihtiadi}} (\bibinfo {collaboration} {COSINE-100 Collaboration}),\ }\href {\doibase 10.22323/1.301.0883} {\bibfield  {journal} {\bibinfo  {journal} {Proc. Sci.}\ }\textbf {\bibinfo {volume} {ICRC2017}},\ \bibinfo {pages} {883} (\bibinfo {year} {2018})}\BibitemShut {NoStop}%
\bibitem [{\citenamefont {Ruddick}(1996)}]{ruddick1996underground}%
  \BibitemOpen
  \bibfield  {author} {\bibinfo {author} {\bibfnamefont {K.}~\bibnamefont {Ruddick}},\ }\href@noop {} {\bibfield  {journal} {\bibinfo  {journal} {MINOS internal note NuMI-L-210}\ } (\bibinfo {year} {1996})}\BibitemShut {NoStop}%
\bibitem [{\citenamefont {Abe}\ \emph {et~al.}(2018)\citenamefont {Abe} \emph {et~al.}}]{Hyper-Kamiokande:2018ofw}%
  \BibitemOpen
  \bibfield  {author} {\bibinfo {author} {\bibfnamefont {K.}~\bibnamefont {Abe}} \emph {et~al.} (\bibinfo {collaboration} {Hyper-Kamiokande Collaboration}),\ }\href@noop {} {\  (\bibinfo {year} {2018})},\ \Eprint {http://arxiv.org/abs/1805.04163} {arXiv:1805.04163 [physics.ins-det]} \BibitemShut {NoStop}%
\bibitem [{\citenamefont {Robinson}\ \emph {et~al.}(2003)\citenamefont {Robinson}, \citenamefont {Kudryavtsev}, \citenamefont {Luscher}, \citenamefont {McMillan}, \citenamefont {Lightfoot}, \citenamefont {Spooner}, \citenamefont {Smith},\ and\ \citenamefont {Liubarsky}}]{Robinson:2003zj}%
  \BibitemOpen
  \bibfield  {author} {\bibinfo {author} {\bibfnamefont {M.}~\bibnamefont {Robinson}}, \bibinfo {author} {\bibfnamefont {V.~A.}\ \bibnamefont {Kudryavtsev}}, \bibinfo {author} {\bibfnamefont {R.}~\bibnamefont {Luscher}}, \bibinfo {author} {\bibfnamefont {J.~E.}\ \bibnamefont {McMillan}}, \bibinfo {author} {\bibfnamefont {P.~K.}\ \bibnamefont {Lightfoot}}, \bibinfo {author} {\bibfnamefont {N.~J.~C.}\ \bibnamefont {Spooner}}, \bibinfo {author} {\bibfnamefont {N.~J.~T.}\ \bibnamefont {Smith}}, \ and\ \bibinfo {author} {\bibfnamefont {I.}~\bibnamefont {Liubarsky}},\ }\href {\doibase 10.1016/S0168-9002(03)01973-9} {\bibfield  {journal} {\bibinfo  {journal} {Nucl. Instrum. Methods Phys. Res., Sect. A}\ }\textbf {\bibinfo {volume} {511}},\ \bibinfo {pages} {347} (\bibinfo {year} {2003})}\BibitemShut {NoStop}%
\bibitem [{\citenamefont {Urquijo}(2016)}]{Urquijo:2016dxd}%
  \BibitemOpen
  \bibfield  {author} {\bibinfo {author} {\bibfnamefont {P.}~\bibnamefont {Urquijo}},\ }\href {\doibase 10.1051/epjconf/201612304002} {\bibfield  {journal} {\bibinfo  {journal} {EPJ Web Conf.}\ }\textbf {\bibinfo {volume} {123}},\ \bibinfo {pages} {04002} (\bibinfo {year} {2016})},\ \Eprint {http://arxiv.org/abs/1605.03299} {1605.03299} \BibitemShut {NoStop}%
\bibitem [{\citenamefont {Barberio}\ \emph {et~al.}(2023)\citenamefont {Barberio} \emph {et~al.}}]{Barberio:2022grv}%
  \BibitemOpen
  \bibfield  {author} {\bibinfo {author} {\bibfnamefont {E.}~\bibnamefont {Barberio}} \emph {et~al.},\ }\href {\doibase 10.22323/1.444.1370} {\bibfield  {journal} {\bibinfo  {journal} {Proc. Sci.}\ }\textbf {\bibinfo {volume} {ICRC2023}},\ \bibinfo {pages} {1370} (\bibinfo {year} {2023})}\BibitemShut {NoStop}%
\bibitem [{\citenamefont {{Capuano}}\ \emph {et~al.}(1998)\citenamefont {{Capuano}}, \citenamefont {{De Luca}}, \citenamefont {{Di Sena}}, \citenamefont {{Gasparini}},\ and\ \citenamefont {{Scarpa}}}]{1998JAG....39...25C}%
  \BibitemOpen
  \bibfield  {author} {\bibinfo {author} {\bibfnamefont {P.}~\bibnamefont {{Capuano}}}, \bibinfo {author} {\bibfnamefont {G.}~\bibnamefont {{De Luca}}}, \bibinfo {author} {\bibfnamefont {F.}~\bibnamefont {{Di Sena}}}, \bibinfo {author} {\bibfnamefont {P.}~\bibnamefont {{Gasparini}}}, \ and\ \bibinfo {author} {\bibfnamefont {R.}~\bibnamefont {{Scarpa}}},\ }\href {\doibase 10.1016/S0926-9851(98)00002-0} {\bibfield  {journal} {\bibinfo  {journal} {J. Appl. Geophys.}\ }\textbf {\bibinfo {volume} {39}},\ \bibinfo {pages} {25} (\bibinfo {year} {1998})}\BibitemShut {NoStop}%
\bibitem [{\citenamefont {Abgrall}\ \emph {et~al.}(2017)\citenamefont {Abgrall} \emph {et~al.}}]{MAJORANA:2016ifg}%
  \BibitemOpen
  \bibfield  {author} {\bibinfo {author} {\bibfnamefont {N.}~\bibnamefont {Abgrall}} \emph {et~al.} (\bibinfo {collaboration} {MAJORANA Collaboration}),\ }\href {\doibase 10.1016/j.astropartphys.2017.01.013} {\bibfield  {journal} {\bibinfo  {journal} {Astropart. Phys.}\ }\textbf {\bibinfo {volume} {93}},\ \bibinfo {pages} {70} (\bibinfo {year} {2017})}\BibitemShut {NoStop}%
\bibitem [{\citenamefont {Heise}(2020)}]{Heise:2017rpu}%
  \BibitemOpen
  \bibfield  {author} {\bibinfo {author} {\bibfnamefont {J.}~\bibnamefont {Heise}},\ }\href {\doibase 10.5281/zenodo.1300395} {\bibfield  {journal} {\bibinfo  {journal} {J. Phys. Conf. Ser.}\ }\textbf {\bibinfo {volume} {1342}},\ \bibinfo {pages} {012085} (\bibinfo {year} {2020})}\BibitemShut {NoStop}%
\bibitem [{\citenamefont {Wu}\ \emph {et~al.}(2013)\citenamefont {Wu} \emph {et~al.}}]{Wu:2013cno}%
  \BibitemOpen
  \bibfield  {author} {\bibinfo {author} {\bibfnamefont {Y.-C.}\ \bibnamefont {Wu}} \emph {et~al.},\ }\href {\doibase 10.1088/1674-1137/37/8/086001} {\bibfield  {journal} {\bibinfo  {journal} {Chin. Phys. C}\ }\textbf {\bibinfo {volume} {37}},\ \bibinfo {pages} {086001} (\bibinfo {year} {2013})}\BibitemShut {NoStop}%
\bibitem [{\citenamefont {Menon}\ and\ \citenamefont {Murthy}(1967)}]{menon1967progress}%
  \BibitemOpen
  \bibfield  {author} {\bibinfo {author} {\bibfnamefont {M.}~\bibnamefont {Menon}}\ and\ \bibinfo {author} {\bibfnamefont {P.~R.}\ \bibnamefont {Murthy}},\ }\enquote {\bibinfo {title} {Cosmic ray intensities deep underground},}\ in\ \href {https://pdg.lbl.gov/2007/AtomicNuclearProperties/standardrock.html} {\emph {\bibinfo {booktitle} {Progress in Elementary Particle and Cosmic Ray Physics}}},\ Vol.~\bibinfo {volume} {9}\ (\bibinfo  {publisher} {North-Holland Publishing Co., Amsterdam},\ \bibinfo {year} {1967})\ pp.\ \bibinfo {pages} {163--243}\BibitemShut {NoStop}%
\bibitem [{\citenamefont {Ambrosio}\ \emph {et~al.}(2003{\natexlab{a}})\citenamefont {Ambrosio} \emph {et~al.}}]{MACRO:2002qsl}%
  \BibitemOpen
  \bibfield  {author} {\bibinfo {author} {\bibfnamefont {M.}~\bibnamefont {Ambrosio}} \emph {et~al.} (\bibinfo {collaboration} {MACRO Collaboration}),\ }\href {\doibase 10.1103/PhysRevD.67.042002} {\bibfield  {journal} {\bibinfo  {journal} {Phys. Rev. D}\ }\textbf {\bibinfo {volume} {67}},\ \bibinfo {pages} {042002} (\bibinfo {year} {2003}{\natexlab{a}})}\BibitemShut {NoStop}%
\bibitem [{\citenamefont {Agostini}\ \emph {et~al.}(2019)\citenamefont {Agostini} \emph {et~al.}}]{Borexino:2018pev}%
  \BibitemOpen
  \bibfield  {author} {\bibinfo {author} {\bibfnamefont {M.}~\bibnamefont {Agostini}} \emph {et~al.} (\bibinfo {collaboration} {Borexino}),\ }\href {\doibase 10.1088/1475-7516/2019/02/046} {\bibfield  {journal} {\bibinfo  {journal} {JCAP}\ }\textbf {\bibinfo {volume} {02}},\ \bibinfo {pages} {046} (\bibinfo {year} {2019})}\BibitemShut {NoStop}%
\bibitem [{\citenamefont {Agafonova}\ \emph {et~al.}(2019)\citenamefont {Agafonova} \emph {et~al.}}]{LVD:2019zlh}%
  \BibitemOpen
  \bibfield  {author} {\bibinfo {author} {\bibfnamefont {N.~Y.}\ \bibnamefont {Agafonova}} \emph {et~al.} (\bibinfo {collaboration} {LVD Collaboration}),\ }\href {\doibase 10.1103/PhysRevD.100.062002} {\bibfield  {journal} {\bibinfo  {journal} {Phys. Rev. D}\ }\textbf {\bibinfo {volume} {100}},\ \bibinfo {pages} {062002} (\bibinfo {year} {2019})}\BibitemShut {NoStop}%
\bibitem [{\citenamefont {{Kokoulin}}\ and\ \citenamefont {{Petrukhin}}(1971)}]{1971ICRC....6.2436K}%
  \BibitemOpen
  \bibfield  {author} {\bibinfo {author} {\bibfnamefont {R.~P.}\ \bibnamefont {{Kokoulin}}}\ and\ \bibinfo {author} {\bibfnamefont {A.~A.}\ \bibnamefont {{Petrukhin}}},\ }in\ \href@noop {} {\emph {\bibinfo {booktitle} {12th International Cosmic Ray Conference (ICRC12)}}},\ Vol.~\bibinfo {volume} {6}\ (\bibinfo  {publisher} {University of Tasmania, Hobart},\ \bibinfo {year} {1971})\ p.\ \bibinfo {pages} {2436}\BibitemShut {NoStop}%
\bibitem [{\citenamefont {Kelner}\ \emph {et~al.}(1995)\citenamefont {Kelner}, \citenamefont {Kokoulin},\ and\ \citenamefont {Petrukhin}}]{Kelner:1995hu}%
  \BibitemOpen
  \bibfield  {author} {\bibinfo {author} {\bibfnamefont {S.~R.}\ \bibnamefont {Kelner}}, \bibinfo {author} {\bibfnamefont {R.~P.}\ \bibnamefont {Kokoulin}}, \ and\ \bibinfo {author} {\bibfnamefont {A.~A.}\ \bibnamefont {Petrukhin}},\ }\href@noop {} {\emph {\bibinfo {title} {About Cross-Section for High-Energy Muon Bremsstrahlung}}},\ \bibinfo {type} {Tech. Rep.}\ \bibinfo {number} {Fprint-95-36}\ (\bibinfo {year} {1995})\BibitemShut {NoStop}%
\bibitem [{\citenamefont {Kelner}\ and\ \citenamefont {Fedotov}(1999)}]{Kelner:1999yf}%
  \BibitemOpen
  \bibfield  {author} {\bibinfo {author} {\bibfnamefont {S.~R.}\ \bibnamefont {Kelner}}\ and\ \bibinfo {author} {\bibfnamefont {A.~M.}\ \bibnamefont {Fedotov}},\ }\href {https://ui.adsabs.harvard.edu/abs/1999PAN....62..272K} {\bibfield  {journal} {\bibinfo  {journal} {Phys. At. Nucl.}\ }\textbf {\bibinfo {volume} {62}},\ \bibinfo {pages} {272} (\bibinfo {year} {1999})}\BibitemShut {NoStop}%
\bibitem [{\citenamefont {Abramowicz}\ \emph {et~al.}(1991)\citenamefont {Abramowicz}, \citenamefont {Levin}, \citenamefont {Levy},\ and\ \citenamefont {Maor}}]{Abramowicz:1991xz}%
  \BibitemOpen
  \bibfield  {author} {\bibinfo {author} {\bibfnamefont {H.}~\bibnamefont {Abramowicz}}, \bibinfo {author} {\bibfnamefont {E.~M.}\ \bibnamefont {Levin}}, \bibinfo {author} {\bibfnamefont {A.}~\bibnamefont {Levy}}, \ and\ \bibinfo {author} {\bibfnamefont {U.}~\bibnamefont {Maor}},\ }\href {\doibase 10.1016/0370-2693(91)90202-2} {\bibfield  {journal} {\bibinfo  {journal} {Phys. Lett. B}\ }\textbf {\bibinfo {volume} {269}},\ \bibinfo {pages} {465} (\bibinfo {year} {1991})}\BibitemShut {NoStop}%
\bibitem [{\citenamefont {Abramowicz}\ and\ \citenamefont {Levy}(1997)}]{Abramowicz:1997ms}%
  \BibitemOpen
  \bibfield  {author} {\bibinfo {author} {\bibfnamefont {H.}~\bibnamefont {Abramowicz}}\ and\ \bibinfo {author} {\bibfnamefont {A.}~\bibnamefont {Levy}},\ }\href@noop {} {\  (\bibinfo {year} {1997})},\ \Eprint {http://arxiv.org/abs/hep-ph/9712415} {arXiv:hep-ph/9712415} \BibitemShut {NoStop}%
\bibitem [{\citenamefont {Rossi}(1952)}]{Rossi:1952kt}%
  \BibitemOpen
  \bibfield  {author} {\bibinfo {author} {\bibfnamefont {B.}~\bibnamefont {Rossi}},\ }\href {https://cds.cern.ch/record/99081} {\emph {\bibinfo {title} {{High Energy Particles}}}},\ Prentice-Hall physics series\ (\bibinfo  {publisher} {{Prentice-Hall, New York}},\ \bibinfo {year} {1952})\BibitemShut {NoStop}%
\bibitem [{\citenamefont {Adamson}\ \emph {et~al.}(2007)\citenamefont {Adamson} \emph {et~al.}}]{MINOS:2007laz}%
  \BibitemOpen
  \bibfield  {author} {\bibinfo {author} {\bibfnamefont {P.}~\bibnamefont {Adamson}} \emph {et~al.} (\bibinfo {collaboration} {MINOS Collaboration}),\ }\href {\doibase 10.1103/PhysRevD.76.052003} {\bibfield  {journal} {\bibinfo  {journal} {Phys. Rev. D}\ }\textbf {\bibinfo {volume} {76}},\ \bibinfo {pages} {052003} (\bibinfo {year} {2007})}\BibitemShut {NoStop}%
\bibitem [{\citenamefont {Mizukoshi}\ \emph {et~al.}(2018)\citenamefont {Mizukoshi} \emph {et~al.}}]{Mizukoshi:2018rnt}%
  \BibitemOpen
  \bibfield  {author} {\bibinfo {author} {\bibfnamefont {K.}~\bibnamefont {Mizukoshi}} \emph {et~al.},\ }\href {\doibase 10.1093/ptep/pty133} {\bibfield  {journal} {\bibinfo  {journal} {Prog. Theor. Exp. Phys.}\ }\textbf {\bibinfo {volume} {2018}},\ \bibinfo {pages} {123C01} (\bibinfo {year} {2018})}\BibitemShut {NoStop}%
\bibitem [{\citenamefont {Chazal}\ \emph {et~al.}(1998)\citenamefont {Chazal}, \citenamefont {Chambon}, \citenamefont {De~Jesus}, \citenamefont {Drain}, \citenamefont {Pastor}, \citenamefont {Vagneron}, \citenamefont {Brissot}, \citenamefont {Cavaignac}, \citenamefont {Stutz},\ and\ \citenamefont {Giraud-Heraud}}]{Chazal:1997qn}%
  \BibitemOpen
  \bibfield  {author} {\bibinfo {author} {\bibfnamefont {V.}~\bibnamefont {Chazal}}, \bibinfo {author} {\bibfnamefont {B.}~\bibnamefont {Chambon}}, \bibinfo {author} {\bibfnamefont {M.}~\bibnamefont {De~Jesus}}, \bibinfo {author} {\bibfnamefont {D.}~\bibnamefont {Drain}}, \bibinfo {author} {\bibfnamefont {C.}~\bibnamefont {Pastor}}, \bibinfo {author} {\bibfnamefont {L.}~\bibnamefont {Vagneron}}, \bibinfo {author} {\bibfnamefont {R.}~\bibnamefont {Brissot}}, \bibinfo {author} {\bibfnamefont {J.~F.}\ \bibnamefont {Cavaignac}}, \bibinfo {author} {\bibfnamefont {A.}~\bibnamefont {Stutz}}, \ and\ \bibinfo {author} {\bibfnamefont {Y.}~\bibnamefont {Giraud-Heraud}},\ }\href {\doibase 10.1016/s0927-6505(98)00012-7} {\bibfield  {journal} {\bibinfo  {journal} {Astropart. Phys.}\ }\textbf {\bibinfo {volume} {9}},\ \bibinfo {pages} {163} (\bibinfo {year} {1998})}\BibitemShut {NoStop}%
\bibitem [{\citenamefont {Rhode}(1993)}]{rhode1993study}%
  \BibitemOpen
  \bibfield  {author} {\bibinfo {author} {\bibfnamefont {W.}~\bibnamefont {Rhode}},\ }\emph {\bibinfo {title} {Study of ultra high energy muons with the Fr{\'e}jus detector}},\ \href@noop {} {Ph.D. thesis} (\bibinfo {year} {1993})\BibitemShut {NoStop}%
\bibitem [{\citenamefont {Kyba}(2006)}]{kyba2006measurement}%
  \BibitemOpen
  \bibfield  {author} {\bibinfo {author} {\bibfnamefont {C.~C.~M.}\ \bibnamefont {Kyba}},\ }\emph {\bibinfo {title} {Measurement of the atmospheric neutrino induced muon flux at the Sudbury Neutrino Observatory}},\ \href@noop {} {Ph.D. thesis} (\bibinfo {year} {2006})\BibitemShut {NoStop}%
\bibitem [{\citenamefont {Ewan}\ \emph {et~al.}(1987)\citenamefont {Ewan} \emph {et~al.}}]{Ewan:1987oqj}%
  \BibitemOpen
  \bibfield  {author} {\bibinfo {author} {\bibfnamefont {G.~T.}\ \bibnamefont {Ewan}} \emph {et~al.},\ }\href@noop {} {\emph {\bibinfo {title} {Sudbury Neutrino Observatory Proposal}}},\ \bibinfo {type} {Tech. Rep.}\ \bibinfo {number} {SNO-87-12}\ (\bibinfo {year} {1987})\BibitemShut {NoStop}%
\bibitem [{\citenamefont {Powers}\ \emph {et~al.}(1978)\citenamefont {Powers}, \citenamefont {Lambert}, \citenamefont {Shaffer}, \citenamefont {Hill},\ and\ \citenamefont {Weart}}]{osti_6441454}%
  \BibitemOpen
  \bibfield  {author} {\bibinfo {author} {\bibfnamefont {D.~W.}\ \bibnamefont {Powers}}, \bibinfo {author} {\bibfnamefont {S.~J.}\ \bibnamefont {Lambert}}, \bibinfo {author} {\bibfnamefont {S.~E.}\ \bibnamefont {Shaffer}}, \bibinfo {author} {\bibfnamefont {L.~R.}\ \bibnamefont {Hill}}, \ and\ \bibinfo {author} {\bibfnamefont {W.~D.}\ \bibnamefont {Weart}},\ }\href {\doibase 10.2172/6441454} {\emph {\bibinfo {title} {Geological Characterisation Report, Waste Isolation Pilot Plant (WIPP) Site, Southeastern New Mexico}}},\ \bibinfo {type} {Tech. Rep.}\ \bibinfo {number} {SAND-78-1596(Vol.1), TRN: 79-004659}\ (\bibinfo {year} {1978})\BibitemShut {NoStop}%
\bibitem [{\citenamefont {Yoon}\ \emph {et~al.}(2021)\citenamefont {Yoon}, \citenamefont {Kim},\ and\ \citenamefont {Park}}]{Yoon:2021tkv}%
  \BibitemOpen
  \bibfield  {author} {\bibinfo {author} {\bibfnamefont {Y.~S.}\ \bibnamefont {Yoon}}, \bibinfo {author} {\bibfnamefont {J.}~\bibnamefont {Kim}}, \ and\ \bibinfo {author} {\bibfnamefont {H.}~\bibnamefont {Park}},\ }\href {\doibase 10.1016/j.astropartphys.2020.102533} {\bibfield  {journal} {\bibinfo  {journal} {Astropart. Phys.}\ }\textbf {\bibinfo {volume} {126}},\ \bibinfo {pages} {102533} (\bibinfo {year} {2021})}\BibitemShut {NoStop}%
\bibitem [{\citenamefont {Mei}\ \emph {et~al.}(2010)\citenamefont {Mei}, \citenamefont {Zhang}, \citenamefont {Thomas},\ and\ \citenamefont {Gray}}]{Mei:2009py}%
  \BibitemOpen
  \bibfield  {author} {\bibinfo {author} {\bibfnamefont {D.-M.}\ \bibnamefont {Mei}}, \bibinfo {author} {\bibfnamefont {C.}~\bibnamefont {Zhang}}, \bibinfo {author} {\bibfnamefont {K.}~\bibnamefont {Thomas}}, \ and\ \bibinfo {author} {\bibfnamefont {F.}~\bibnamefont {Gray}},\ }\href {\doibase 10.1016/j.astropartphys.2010.04.003} {\bibfield  {journal} {\bibinfo  {journal} {Astropart. Phys.}\ }\textbf {\bibinfo {volume} {34}},\ \bibinfo {pages} {33} (\bibinfo {year} {2010})}\BibitemShut {NoStop}%
\bibitem [{\citenamefont {Liu}\ \emph {et~al.}(2022)\citenamefont {Liu} \emph {et~al.}}]{CDEX:2021cll}%
  \BibitemOpen
  \bibfield  {author} {\bibinfo {author} {\bibfnamefont {Z.~Z.}\ \bibnamefont {Liu}} \emph {et~al.} (\bibinfo {collaboration} {Cdex Collaboration}),\ }\href {\doibase 10.1103/PhysRevD.105.052005} {\bibfield  {journal} {\bibinfo  {journal} {Phys. Rev. D}\ }\textbf {\bibinfo {volume} {105}},\ \bibinfo {pages} {052005} (\bibinfo {year} {2022})}\BibitemShut {NoStop}%
\bibitem [{\citenamefont {Dugdale}\ \emph {et~al.}(2010)\citenamefont {Dugdale}, \citenamefont {Wilson}, \citenamefont {Dugdale}, \citenamefont {Funk}, \citenamefont {Bosnjak},\ and\ \citenamefont {Jupp}}]{DUGDALE201041}%
  \BibitemOpen
  \bibfield  {author} {\bibinfo {author} {\bibfnamefont {A.}~\bibnamefont {Dugdale}}, \bibinfo {author} {\bibfnamefont {C.}~\bibnamefont {Wilson}}, \bibinfo {author} {\bibfnamefont {L.}~\bibnamefont {Dugdale}}, \bibinfo {author} {\bibfnamefont {C.}~\bibnamefont {Funk}}, \bibinfo {author} {\bibfnamefont {M.}~\bibnamefont {Bosnjak}}, \ and\ \bibinfo {author} {\bibfnamefont {B.}~\bibnamefont {Jupp}},\ }\href {\doibase https://doi.org/10.1016/j.oregeorev.2009.08.001} {\bibfield  {journal} {\bibinfo  {journal} {Ore Geol. Rev.}\ }\textbf {\bibinfo {volume} {37}},\ \bibinfo {pages} {41} (\bibinfo {year} {2010})}\BibitemShut {NoStop}%
\bibitem [{\citenamefont {Zurowski}(2023)}]{Zurowski:2023wwx}%
  \BibitemOpen
  \bibfield  {author} {\bibinfo {author} {\bibfnamefont {M.~J.}\ \bibnamefont {Zurowski}} (\bibinfo {collaboration} {SABRE South Collaboration}),\ }\href {\doibase 10.21468/SciPostPhysProc.12.029} {\bibfield  {journal} {\bibinfo  {journal} {SciPost Phys. Proc.}\ }\textbf {\bibinfo {volume} {12}},\ \bibinfo {pages} {029} (\bibinfo {year} {2023})}\BibitemShut {NoStop}%
\bibitem [{\citenamefont {Sternheimer}(1952)}]{Sternheimer:1952jn}%
  \BibitemOpen
  \bibfield  {author} {\bibinfo {author} {\bibfnamefont {R.~M.}\ \bibnamefont {Sternheimer}},\ }\href {\doibase 10.1103/PhysRev.88.851} {\bibfield  {journal} {\bibinfo  {journal} {Phys. Rev.}\ }\textbf {\bibinfo {volume} {88}},\ \bibinfo {pages} {851} (\bibinfo {year} {1952})}\BibitemShut {NoStop}%
\bibitem [{\citenamefont {Sternheimer}\ \emph {et~al.}(1984)\citenamefont {Sternheimer}, \citenamefont {Berger},\ and\ \citenamefont {Seltzer}}]{Sternheimer:1983mb}%
  \BibitemOpen
  \bibfield  {author} {\bibinfo {author} {\bibfnamefont {R.~M.}\ \bibnamefont {Sternheimer}}, \bibinfo {author} {\bibfnamefont {M.~J.}\ \bibnamefont {Berger}}, \ and\ \bibinfo {author} {\bibfnamefont {S.~M.}\ \bibnamefont {Seltzer}},\ }\href {\doibase 10.1016/0092-640X(84)90002-0} {\bibfield  {journal} {\bibinfo  {journal} {At. Data Nucl. Data Tables}\ }\textbf {\bibinfo {volume} {30}},\ \bibinfo {pages} {261} (\bibinfo {year} {1984})}\BibitemShut {NoStop}%
\bibitem [{\citenamefont {Lohmann}\ \emph {et~al.}(1985)\citenamefont {Lohmann}, \citenamefont {Kopp},\ and\ \citenamefont {Voss}}]{lohmann1985energy}%
  \BibitemOpen
  \bibfield  {author} {\bibinfo {author} {\bibfnamefont {W.}~\bibnamefont {Lohmann}}, \bibinfo {author} {\bibfnamefont {R.}~\bibnamefont {Kopp}}, \ and\ \bibinfo {author} {\bibfnamefont {R.}~\bibnamefont {Voss}},\ }\href@noop {} {\emph {\bibinfo {title} {Energy Loss of Muons in the Energy Range 1--10000 GeV}}},\ \bibinfo {type} {Tech. Rep.}\ \bibinfo {number} {CERN-85-03, CERN-YELLOW-85-03}\ (\bibinfo  {institution} {European Organisation for Nuclear Research},\ \bibinfo {year} {1985})\BibitemShut {NoStop}%
\bibitem [{\citenamefont {Groom}\ \emph {et~al.}(2001)\citenamefont {Groom}, \citenamefont {Mokhov},\ and\ \citenamefont {Striganov}}]{Groom:2001kq}%
  \BibitemOpen
  \bibfield  {author} {\bibinfo {author} {\bibfnamefont {D.~E.}\ \bibnamefont {Groom}}, \bibinfo {author} {\bibfnamefont {N.~V.}\ \bibnamefont {Mokhov}}, \ and\ \bibinfo {author} {\bibfnamefont {S.~I.}\ \bibnamefont {Striganov}},\ }\href {\doibase 10.1006/adnd.2001.0861} {\bibfield  {journal} {\bibinfo  {journal} {At. Data Nucl. Data Tables}\ }\textbf {\bibinfo {volume} {78}},\ \bibinfo {pages} {183} (\bibinfo {year} {2001})}\BibitemShut {NoStop}%
\bibitem [{\citenamefont {on~Extension to~the Standard~Atmosphere}(1976)}]{atmosphere1976national}%
  \BibitemOpen
  \bibfield  {author} {\bibinfo {author} {\bibfnamefont {U.~S.~C.}\ \bibnamefont {on~Extension to~the Standard~Atmosphere}},\ }\href {https://www.ngdc.noaa.gov/stp/space-weather/online-publications/miscellaneous/us-standard-atmosphere-1976/us-standard-atmosphere_st76-1562_noaa.pdf} {\emph {\bibinfo {title} {U.S. Standard Atmosphere, 1976}}},\ ADA 035 728\ (\bibinfo  {publisher} {National Oceanic and Atmospheric Administration},\ \bibinfo {year} {1976})\BibitemShut {NoStop}%
\bibitem [{\citenamefont {Prihtiadi}\ \emph {et~al.}(2021)\citenamefont {Prihtiadi} \emph {et~al.}}]{COSINE-100:2020jml}%
  \BibitemOpen
  \bibfield  {author} {\bibinfo {author} {\bibfnamefont {H.}~\bibnamefont {Prihtiadi}} \emph {et~al.} (\bibinfo {collaboration} {COSINE-100 Collaboration}),\ }\href {\doibase 10.1088/1475-7516/2021/02/013} {\bibfield  {journal} {\bibinfo  {journal} {J. Cosmol. Astropart. Phys.}\ }\textbf {\bibinfo {volume} {02}},\ \bibinfo {pages} {013} (\bibinfo {year} {2021})}\BibitemShut {NoStop}%
\bibitem [{\citenamefont {Zhang}\ and\ \citenamefont {Mei}(2014)}]{Zhang:2014jsq}%
  \BibitemOpen
  \bibfield  {author} {\bibinfo {author} {\bibfnamefont {C.}~\bibnamefont {Zhang}}\ and\ \bibinfo {author} {\bibfnamefont {D.~M.}\ \bibnamefont {Mei}},\ }\href {\doibase 10.1103/PhysRevD.90.122003} {\bibfield  {journal} {\bibinfo  {journal} {Phys. Rev. D}\ }\textbf {\bibinfo {volume} {90}},\ \bibinfo {pages} {122003} (\bibinfo {year} {2014})}\BibitemShut {NoStop}%
\bibitem [{\citenamefont {Melbourne}(2021)}]{Melbourne:2021wxo}%
  \BibitemOpen
  \bibfield  {author} {\bibinfo {author} {\bibfnamefont {W.~D.}\ \bibnamefont {Melbourne}},\ }\href {\doibase 10.1088/1742-6596/2156/1/012064} {\bibfield  {journal} {\bibinfo  {journal} {J. Phys. Conf. Ser.}\ }\textbf {\bibinfo {volume} {2156}},\ \bibinfo {pages} {012064} (\bibinfo {year} {2021})}\BibitemShut {NoStop}%
\bibitem [{\citenamefont {Schmidt}\ \emph {et~al.}(2013)\citenamefont {Schmidt} \emph {et~al.}}]{EDELWEISS:2013kzp}%
  \BibitemOpen
  \bibfield  {author} {\bibinfo {author} {\bibfnamefont {B.}~\bibnamefont {Schmidt}} \emph {et~al.} (\bibinfo {collaboration} {EDELWEISS Collaboration}),\ }\href {\doibase 10.1016/j.astropartphys.2013.01.014} {\bibfield  {journal} {\bibinfo  {journal} {Astropart. Phys.}\ }\textbf {\bibinfo {volume} {44}},\ \bibinfo {pages} {28} (\bibinfo {year} {2013})}\BibitemShut {NoStop}%
\bibitem [{\citenamefont {Cherry}\ \emph {et~al.}(1983)\citenamefont {Cherry}, \citenamefont {Deakyne}, \citenamefont {Lande}, \citenamefont {Lee}, \citenamefont {Steinberg}, \citenamefont {Cleveland},\ and\ \citenamefont {Fenyves}}]{Cherry:1983dp}%
  \BibitemOpen
  \bibfield  {author} {\bibinfo {author} {\bibfnamefont {M.~L.}\ \bibnamefont {Cherry}}, \bibinfo {author} {\bibfnamefont {M.}~\bibnamefont {Deakyne}}, \bibinfo {author} {\bibfnamefont {K.}~\bibnamefont {Lande}}, \bibinfo {author} {\bibfnamefont {C.~K.}\ \bibnamefont {Lee}}, \bibinfo {author} {\bibfnamefont {R.~I.}\ \bibnamefont {Steinberg}}, \bibinfo {author} {\bibfnamefont {B.~T.}\ \bibnamefont {Cleveland}}, \ and\ \bibinfo {author} {\bibfnamefont {E.~J.}\ \bibnamefont {Fenyves}},\ }\href {\doibase 10.1103/PhysRevD.27.1444} {\bibfield  {journal} {\bibinfo  {journal} {Phys. Rev. D}\ }\textbf {\bibinfo {volume} {27}},\ \bibinfo {pages} {1444} (\bibinfo {year} {1983})}\BibitemShut {NoStop}%
\bibitem [{\citenamefont {Ihm}(2018)}]{ihm2018through}%
  \BibitemOpen
  \bibfield  {author} {\bibinfo {author} {\bibfnamefont {C.~M.}\ \bibnamefont {Ihm}},\ }\emph {\bibinfo {title} {Through-going Muons in the LUX Dark Matter Search}},\ \href {https://escholarship.org/uc/item/3t45p4pv} {Ph.D. thesis},\ \bibinfo  {school} {UC Berkeley} (\bibinfo {year} {2018})\BibitemShut {NoStop}%
\bibitem [{\citenamefont {Wan}(2019)}]{Wan:2019qml}%
  \BibitemOpen
  \bibfield  {author} {\bibinfo {author} {\bibfnamefont {L.}~\bibnamefont {Wan}} (\bibinfo {collaboration} {Jinping Collaboration}),\ }\enquote {\bibinfo {title} {{Simulation and sensitivity studies for solar neutrinos at Jinping}},}\ in\ \href {\doibase 10.1142/9789811204296_0023} {\emph {\bibinfo {booktitle} {{Proceedings, 5th International Solar Neutrino Conference}: {Dresden, Germany, 2018}}}},\ \bibinfo {editor} {edited by\ \bibinfo {editor} {\bibfnamefont {M.}~\bibnamefont {Meyer}}\ and\ \bibinfo {editor} {\bibfnamefont {K.}~\bibnamefont {Zuber}}}\ (\bibinfo  {publisher} {Wold Scientific, Dresden},\ \bibinfo {year} {2019})\ pp.\ \bibinfo {pages} {381--389}\BibitemShut {NoStop}%
\bibitem [{\citenamefont {Woodley}\ \emph {et~al.}(2023)\citenamefont {Woodley}, \citenamefont {Fedynitch},\ and\ \citenamefont {Piro}}]{Woodley:2023gnf}%
  \BibitemOpen
  \bibfield  {author} {\bibinfo {author} {\bibfnamefont {W.}~\bibnamefont {Woodley}}, \bibinfo {author} {\bibfnamefont {A.}~\bibnamefont {Fedynitch}}, \ and\ \bibinfo {author} {\bibfnamefont {M.-C.}\ \bibnamefont {Piro}},\ }\href {\doibase 10.22323/1.444.0476} {\bibfield  {journal} {\bibinfo  {journal} {Proc. Sci.}\ }\textbf {\bibinfo {volume} {ICRC2023}},\ \bibinfo {pages} {476} (\bibinfo {year} {2023})}\BibitemShut {NoStop}%
\bibitem [{\citenamefont {Honda}\ \emph {et~al.}(2019)\citenamefont {Honda}, \citenamefont {Sajjad~Athar}, \citenamefont {Kajita}, \citenamefont {Kasahara},\ and\ \citenamefont {Midorikawa}}]{Honda:2019ymh}%
  \BibitemOpen
  \bibfield  {author} {\bibinfo {author} {\bibfnamefont {M.}~\bibnamefont {Honda}}, \bibinfo {author} {\bibfnamefont {M.}~\bibnamefont {Sajjad~Athar}}, \bibinfo {author} {\bibfnamefont {T.}~\bibnamefont {Kajita}}, \bibinfo {author} {\bibfnamefont {K.}~\bibnamefont {Kasahara}}, \ and\ \bibinfo {author} {\bibfnamefont {S.}~\bibnamefont {Midorikawa}},\ }\href {\doibase 10.1103/PhysRevD.100.123022} {\bibfield  {journal} {\bibinfo  {journal} {Phys. Rev. D}\ }\textbf {\bibinfo {volume} {100}},\ \bibinfo {pages} {123022} (\bibinfo {year} {2019})}\BibitemShut {NoStop}%
\bibitem [{\citenamefont {Prihtiadi}\ \emph {et~al.}(2018)\citenamefont {Prihtiadi} \emph {et~al.}}]{COSINE-100:2017dsl}%
  \BibitemOpen
  \bibfield  {author} {\bibinfo {author} {\bibfnamefont {H.}~\bibnamefont {Prihtiadi}} \emph {et~al.} (\bibinfo {collaboration} {COSINE-100 Collaboration}),\ }\href {\doibase 10.1088/1748-0221/13/02/T02007} {\bibfield  {journal} {\bibinfo  {journal} {J. Instrum.}\ }\textbf {\bibinfo {volume} {13}},\ \bibinfo {pages} {T02007} (\bibinfo {year} {2018})}\BibitemShut {NoStop}%
\bibitem [{\citenamefont {Jing-Jun}\ \emph {et~al.}(2005)\citenamefont {Jing-Jun}, \citenamefont {Ke-Jun}, \citenamefont {Yuan-Jing},\ and\ \citenamefont {Jin}}]{kims2005}%
  \BibitemOpen
  \bibfield  {author} {\bibinfo {author} {\bibfnamefont {Z.}~\bibnamefont {Jing-Jun}}, \bibinfo {author} {\bibfnamefont {K.}~\bibnamefont {Ke-Jun}}, \bibinfo {author} {\bibfnamefont {L.}~\bibnamefont {Yuan-Jing}}, \ and\ \bibinfo {author} {\bibfnamefont {L.}~\bibnamefont {Jin}},\ }\href {http://hepnp.ihep.ac.cn/en/article/id/78c8e154-b0df-4017-9f07-3046b3ef9921} {\bibfield  {journal} {\bibinfo  {journal} {Chin. Phys. C}\ }\textbf {\bibinfo {volume} {29}},\ \bibinfo {pages} {721} (\bibinfo {year} {2005})}\BibitemShut {NoStop}%
\bibitem [{\citenamefont {Lee}\ \emph {et~al.}(2006)\citenamefont {Lee} \emph {et~al.}}]{Kims:2005dol}%
  \BibitemOpen
  \bibfield  {author} {\bibinfo {author} {\bibfnamefont {H.~S.}\ \bibnamefont {Lee}} \emph {et~al.} (\bibinfo {collaboration} {Kims Collaboration}),\ }\href {\doibase 10.1016/j.physletb.2005.12.035} {\bibfield  {journal} {\bibinfo  {journal} {Phys. Lett. B}\ }\textbf {\bibinfo {volume} {633}},\ \bibinfo {pages} {201} (\bibinfo {year} {2006})}\BibitemShut {NoStop}%
\bibitem [{\citenamefont {Pronost}\ \emph {et~al.}(2018)\citenamefont {Pronost}, \citenamefont {Ikeda}, \citenamefont {Nakamura}, \citenamefont {Sekiya},\ and\ \citenamefont {Tasaka}}]{Pronost:2018ghn}%
  \BibitemOpen
  \bibfield  {author} {\bibinfo {author} {\bibfnamefont {G.}~\bibnamefont {Pronost}}, \bibinfo {author} {\bibfnamefont {M.}~\bibnamefont {Ikeda}}, \bibinfo {author} {\bibfnamefont {T.}~\bibnamefont {Nakamura}}, \bibinfo {author} {\bibfnamefont {H.}~\bibnamefont {Sekiya}}, \ and\ \bibinfo {author} {\bibfnamefont {S.}~\bibnamefont {Tasaka}},\ }\href {\doibase 10.1093/ptep/pty091} {\bibfield  {journal} {\bibinfo  {journal} {Prog. Theor. Exp. Phys.}\ }\textbf {\bibinfo {volume} {2018}},\ \bibinfo {pages} {093H01} (\bibinfo {year} {2018})}\BibitemShut {NoStop}%
\bibitem [{\citenamefont {Kamat}(2005)}]{Kamat:2005ct}%
  \BibitemOpen
  \bibfield  {author} {\bibinfo {author} {\bibfnamefont {S.}~\bibnamefont {Kamat}},\ }\emph {\bibinfo {title} {{Extending the Sensitivity to the Detection of WIMP Dark Matter with an Improved Understanding of the Limiting Neutron Backgrounds}}},\ \href {\doibase 10.2172/15017230} {Ph.D. thesis},\ \bibinfo  {school} {Case Western Reserve U.} (\bibinfo {year} {2005})\BibitemShut {NoStop}%
\bibitem [{\citenamefont {Adamson}\ \emph {et~al.}(2015)\citenamefont {Adamson} \emph {et~al.}}]{MINOS:2015ikk}%
  \BibitemOpen
  \bibfield  {author} {\bibinfo {author} {\bibfnamefont {P.}~\bibnamefont {Adamson}} \emph {et~al.} (\bibinfo {collaboration} {MINOS Collaboration}),\ }\href {\doibase 10.1103/PhysRevD.91.112006} {\bibfield  {journal} {\bibinfo  {journal} {Phys. Rev. D}\ }\textbf {\bibinfo {volume} {91}},\ \bibinfo {pages} {112006} (\bibinfo {year} {2015})}\BibitemShut {NoStop}%
\bibitem [{\citenamefont {Akerib}\ \emph {et~al.}(2005)\citenamefont {Akerib} \emph {et~al.}}]{CDMS:2005jsf}%
  \BibitemOpen
  \bibfield  {author} {\bibinfo {author} {\bibfnamefont {D.~S.}\ \bibnamefont {Akerib}} \emph {et~al.} (\bibinfo {collaboration} {CDMS Collaboration}),\ }\href {\doibase 10.1103/PhysRevD.72.052009} {\bibfield  {journal} {\bibinfo  {journal} {Phys. Rev. D}\ }\textbf {\bibinfo {volume} {72}},\ \bibinfo {pages} {052009} (\bibinfo {year} {2005})}\BibitemShut {NoStop}%
\bibitem [{\citenamefont {Adamson}\ \emph {et~al.}(2010)\citenamefont {Adamson} \emph {et~al.}}]{MINOS:2009njg}%
  \BibitemOpen
  \bibfield  {author} {\bibinfo {author} {\bibfnamefont {P.}~\bibnamefont {Adamson}} \emph {et~al.} (\bibinfo {collaboration} {MINOS Collaboration}),\ }\href {\doibase 10.1103/PhysRevD.81.012001} {\bibfield  {journal} {\bibinfo  {journal} {Phys. Rev. D}\ }\textbf {\bibinfo {volume} {81}},\ \bibinfo {pages} {012001} (\bibinfo {year} {2010})}\BibitemShut {NoStop}%
\bibitem [{\citenamefont {Adamson}\ \emph {et~al.}(2020)\citenamefont {Adamson} \emph {et~al.}}]{MINOS:2020iqj}%
  \BibitemOpen
  \bibfield  {author} {\bibinfo {author} {\bibfnamefont {P.}~\bibnamefont {Adamson}} \emph {et~al.} (\bibinfo {collaboration} {MINOS+, Daya Bay Collaborations}),\ }\href {\doibase 10.1103/PhysRevLett.125.071801} {\bibfield  {journal} {\bibinfo  {journal} {Phys. Rev. Lett.}\ }\textbf {\bibinfo {volume} {125}},\ \bibinfo {pages} {071801} (\bibinfo {year} {2020})}\BibitemShut {NoStop}%
\bibitem [{\citenamefont {Kasahara}\ \emph {et~al.}(1997)\citenamefont {Kasahara} \emph {et~al.}}]{Soudan-2:1996hoz}%
  \BibitemOpen
  \bibfield  {author} {\bibinfo {author} {\bibfnamefont {S.~M.}\ \bibnamefont {Kasahara}} \emph {et~al.} (\bibinfo {collaboration} {Soudan-2 Collaboration}),\ }\href {\doibase 10.1103/PhysRevD.55.5282} {\bibfield  {journal} {\bibinfo  {journal} {Phys. Rev. D}\ }\textbf {\bibinfo {volume} {55}},\ \bibinfo {pages} {5282} (\bibinfo {year} {1997})}\BibitemShut {NoStop}%
\bibitem [{\citenamefont {Caddey}\ \emph {et~al.}(1991)\citenamefont {Caddey} \emph {et~al.}}]{caddey1991homestake}%
  \BibitemOpen
  \bibfield  {author} {\bibinfo {author} {\bibfnamefont {S.~W.}\ \bibnamefont {Caddey}} \emph {et~al.},\ }\href {\doibase 10.3133/b1857J} {\emph {\bibinfo {title} {The Homestake Gold Mine, an Early Proterozoic Iron-Formation-Hosted Gold Deposit, Lawrence County, South Dakota}}},\ \bibinfo {number} {1857}\ (\bibinfo  {publisher} {US Government Printing Office, Ann Arbor},\ \bibinfo {year} {1991})\BibitemShut {NoStop}%
\bibitem [{\citenamefont {Hart}\ \emph {et~al.}(2014)\citenamefont {Hart}, \citenamefont {Trancynger}, \citenamefont {Roggenthen},\ and\ \citenamefont {Heise}}]{hart2014topographic}%
  \BibitemOpen
  \bibfield  {author} {\bibinfo {author} {\bibfnamefont {K.}~\bibnamefont {Hart}}, \bibinfo {author} {\bibfnamefont {T.}~\bibnamefont {Trancynger}}, \bibinfo {author} {\bibfnamefont {W.}~\bibnamefont {Roggenthen}}, \ and\ \bibinfo {author} {\bibfnamefont {J.}~\bibnamefont {Heise}},\ }in\ \href@noop {} {\emph {\bibinfo {booktitle} {Proceedings of the South Dakota Academy of Science}}},\ Vol.~\bibinfo {volume} {93}\ (\bibinfo  {publisher} {South Dakota Academy of Science, Vermillion},\ \bibinfo {year} {2014})\BibitemShut {NoStop}%
\bibitem [{\citenamefont {Ageron}\ \emph {et~al.}(2020)\citenamefont {Ageron} \emph {et~al.}}]{KM3NeT:2019jfa}%
  \BibitemOpen
  \bibfield  {author} {\bibinfo {author} {\bibfnamefont {M.}~\bibnamefont {Ageron}} \emph {et~al.} (\bibinfo {collaboration} {KM3NeT Collaboration}),\ }\href {\doibase 10.1140/epjc/s10052-020-7629-z} {\bibfield  {journal} {\bibinfo  {journal} {Eur. Phys. J. C}\ }\textbf {\bibinfo {volume} {80}},\ \bibinfo {pages} {99} (\bibinfo {year} {2020})}\BibitemShut {NoStop}%
\bibitem [{\citenamefont {Aiello}\ \emph {et~al.}(2024)\citenamefont {Aiello} \emph {et~al.}}]{KM3NeT:2024buf}%
  \BibitemOpen
  \bibfield  {author} {\bibinfo {author} {\bibfnamefont {S.}~\bibnamefont {Aiello}} \emph {et~al.} (\bibinfo {collaboration} {KM3NeT Collaboration}),\ }\href@noop {} {\bibfield  {journal} {\bibinfo  {journal} {Eur. Phys. J. C}\ }\textbf {\bibinfo {volume} {84}},\ \bibinfo {pages} {696} (\bibinfo {year} {2024})}\BibitemShut {NoStop}%
\bibitem [{\citenamefont {Tang}\ \emph {et~al.}(2006)\citenamefont {Tang}, \citenamefont {Horton-Smith}, \citenamefont {Kudryavtsev},\ and\ \citenamefont {Tonazzo}}]{Tang:2006uu}%
  \BibitemOpen
  \bibfield  {author} {\bibinfo {author} {\bibfnamefont {A.}~\bibnamefont {Tang}}, \bibinfo {author} {\bibfnamefont {G.}~\bibnamefont {Horton-Smith}}, \bibinfo {author} {\bibfnamefont {V.~A.}\ \bibnamefont {Kudryavtsev}}, \ and\ \bibinfo {author} {\bibfnamefont {A.}~\bibnamefont {Tonazzo}},\ }\href {\doibase 10.1103/PhysRevD.74.053007} {\bibfield  {journal} {\bibinfo  {journal} {Phys. Rev. D}\ }\textbf {\bibinfo {volume} {74}},\ \bibinfo {pages} {053007} (\bibinfo {year} {2006})}\BibitemShut {NoStop}%
\bibitem [{\citenamefont {Gaisser}\ \emph {et~al.}(2016)\citenamefont {Gaisser}, \citenamefont {Engel},\ and\ \citenamefont {Resconi}}]{Gaisser:2016uoy}%
  \BibitemOpen
  \bibfield  {author} {\bibinfo {author} {\bibfnamefont {T.~K.}\ \bibnamefont {Gaisser}}, \bibinfo {author} {\bibfnamefont {R.}~\bibnamefont {Engel}}, \ and\ \bibinfo {author} {\bibfnamefont {E.}~\bibnamefont {Resconi}},\ }\href@noop {} {\emph {\bibinfo {title} {{Cosmic Rays and Particle Physics}: {2nd Edition}}}}\ (\bibinfo  {publisher} {Cambridge University Press, Cambridge, England},\ \bibinfo {year} {2016})\BibitemShut {NoStop}%
\bibitem [{\citenamefont {Battistoni}\ \emph {et~al.}(1999)\citenamefont {Battistoni} \emph {et~al.}}]{MACRO:1998zgn}%
  \BibitemOpen
  \bibfield  {author} {\bibinfo {author} {\bibfnamefont {G.}~\bibnamefont {Battistoni}} \emph {et~al.} (\bibinfo {collaboration} {MACRO Collaboration}),\ }\href@noop {} {\bibfield  {journal} {\bibinfo  {journal} {Italian Phys. Soc. Proc.}\ }\textbf {\bibinfo {volume} {65}},\ \bibinfo {pages} {419} (\bibinfo {year} {1999})}\BibitemShut {NoStop}%
\bibitem [{\citenamefont {Ambrosio}\ \emph {et~al.}(2003{\natexlab{b}})\citenamefont {Ambrosio} \emph {et~al.}}]{MACRO:2002jmi}%
  \BibitemOpen
  \bibfield  {author} {\bibinfo {author} {\bibfnamefont {M.}~\bibnamefont {Ambrosio}} \emph {et~al.} (\bibinfo {collaboration} {MACRO Collaboration}),\ }\href {\doibase 10.1016/S0927-6505(02)00217-7} {\bibfield  {journal} {\bibinfo  {journal} {Astropart. Phys.}\ }\textbf {\bibinfo {volume} {19}},\ \bibinfo {pages} {313} (\bibinfo {year} {2003}{\natexlab{b}})}\BibitemShut {NoStop}%
\bibitem [{\citenamefont {Bussino}\ \emph {et~al.}(1994)\citenamefont {Bussino}, \citenamefont {Chiera}, \citenamefont {Lamanna}, \citenamefont {Bilokon},\ and\ \citenamefont {Miller}}]{Bussino:1994br}%
  \BibitemOpen
  \bibfield  {author} {\bibinfo {author} {\bibfnamefont {S.}~\bibnamefont {Bussino}}, \bibinfo {author} {\bibfnamefont {C.}~\bibnamefont {Chiera}}, \bibinfo {author} {\bibfnamefont {E.}~\bibnamefont {Lamanna}}, \bibinfo {author} {\bibfnamefont {H.}~\bibnamefont {Bilokon}}, \ and\ \bibinfo {author} {\bibfnamefont {L.}~\bibnamefont {Miller}},\ }\href@noop {} {\emph {\bibinfo {title} {Muon Survival Probabilities in Gran Sasso Rock}}},\ \bibinfo {type} {Tech. Rep.}\ \bibinfo {number} {Lngs-94-92}\ (\bibinfo {year} {1994})\BibitemShut {NoStop}%
\bibitem [{\citenamefont {Wulandari}\ \emph {et~al.}(2004)\citenamefont {Wulandari}, \citenamefont {Jochum}, \citenamefont {Rau},\ and\ \citenamefont {von Feilitzsch}}]{Wulandari:2004bj}%
  \BibitemOpen
  \bibfield  {author} {\bibinfo {author} {\bibfnamefont {H.}~\bibnamefont {Wulandari}}, \bibinfo {author} {\bibfnamefont {J.}~\bibnamefont {Jochum}}, \bibinfo {author} {\bibfnamefont {W.}~\bibnamefont {Rau}}, \ and\ \bibinfo {author} {\bibfnamefont {F.}~\bibnamefont {von Feilitzsch}},\ }\href@noop {} {\  (\bibinfo {year} {2004})},\ \Eprint {http://arxiv.org/abs/hep-ex/0401032} {arXiv:hep-ex/0401032} \BibitemShut {NoStop}%
\bibitem [{\citenamefont {Riehn}\ \emph {et~al.}(2020)\citenamefont {Riehn}, \citenamefont {Engel}, \citenamefont {Fedynitch}, \citenamefont {Gaisser},\ and\ \citenamefont {Stanev}}]{Riehn:2019jet}%
  \BibitemOpen
  \bibfield  {author} {\bibinfo {author} {\bibfnamefont {F.}~\bibnamefont {Riehn}}, \bibinfo {author} {\bibfnamefont {R.}~\bibnamefont {Engel}}, \bibinfo {author} {\bibfnamefont {A.}~\bibnamefont {Fedynitch}}, \bibinfo {author} {\bibfnamefont {T.~K.}\ \bibnamefont {Gaisser}}, \ and\ \bibinfo {author} {\bibfnamefont {T.}~\bibnamefont {Stanev}},\ }\href {\doibase 10.1103/PhysRevD.102.063002} {\bibfield  {journal} {\bibinfo  {journal} {Phys. Rev. D}\ }\textbf {\bibinfo {volume} {102}},\ \bibinfo {pages} {063002} (\bibinfo {year} {2020})}\BibitemShut {NoStop}%
\bibitem [{\citenamefont {Gaisser}(2012)}]{Gaisser:2011klf}%
  \BibitemOpen
  \bibfield  {author} {\bibinfo {author} {\bibfnamefont {T.~K.}\ \bibnamefont {Gaisser}},\ }\href {\doibase 10.1016/j.astropartphys.2012.02.010} {\bibfield  {journal} {\bibinfo  {journal} {Astropart. Phys.}\ }\textbf {\bibinfo {volume} {35}},\ \bibinfo {pages} {801} (\bibinfo {year} {2012})}\BibitemShut {NoStop}%
\end{thebibliography}%

\end{document}